\documentclass[fleqn,usenatbib]{mnras}

\usepackage{newtxtext,newtxmath}

\usepackage[T1]{fontenc}
\usepackage{ae,aecompl}

\usepackage{graphicx}	
\usepackage{amsmath}	
\usepackage{amsbsy}	
\usepackage{pifont}
\usepackage{makecell}
\usepackage{booktabs}
\usepackage{multirow}
\usepackage{mathtools}
\usepackage{xcolor}
\usepackage{newtxtext,newtxmath}  
\usepackage{float}

\defcitealias{Owen2021MNRAS}{O21}

\title[EGB imprints from the physical properties of SFGs]{The extragalactic $\gamma$-ray background: imprints from the physical properties and evolution of star-forming galaxy populations
}

\author[Owen, Kong \& Lee]{
Ellis R. Owen$^{1,2}$\thanks{E-mail: erowen@gapp.nthu.edu.tw (ERO)}, 
Albert K. H. Kong$^{1}$, Khee-Gan Lee$^{3}$
\\
$^{1}$Institute of Astronomy, National Tsing Hua University, Hsinchu, Taiwan (ROC) \\
$^{2}$Center for Informatics and Computation in Astronomy, National Tsing Hua University, Hsinchu, Taiwan (ROC)\\
$^{3}$Kavli Institute for the Physics and Mathematics of the Universe, The University of Tokyo, Kashiwa, Chiba 277-8583, Japan
}

\date{Accepted XXX. Received YYY; in original form ZZZ}

\pubyear{2020}

\begin{document}
\label{firstpage}
\pagerange{\pageref{firstpage}--\pageref{lastpage}}
\maketitle

\begin{abstract}
Star-forming galaxies (SFGs) harbour an abundant reservoir of cosmic rays (CRs). At GeV energies, these CRs undergo interactions with their environment to produce $\gamma$-rays, and the unresolved $\gamma$-ray emission from populations of SFGs form a component of the {isotropic} extragalactic $\gamma$-ray background (EGB). In this work, we investigate the contribution to the 0.01 -- 50 GeV EGB from SFG populations located up to redshift $z=3$. We find this is dominated by starbursts, while the contribution from 
main sequence SFGs is marginal at all energies. We also demonstrate that most of the $\gamma$-ray contribution from SFGs emanates from low mass galaxies, with over 80 per cent of the emission originating from galaxies with stellar masses below $10^8 \;\!{\rm M}_{\odot}$. {Many of these galaxies are located at relatively high redshift, with their peak EGB contribution arising $\sim$ 700 Myr before the noon of cosmic star-formation. We find that the precise redshift distributions of EGB sources at different energies imprint intensity signatures at different angular scales, which may allow their contribution to be distinguished using analyses of small-scale EGB intensity anisotropies, particularly if the diffuse EGB is dominated by hadronic CR-driven $\gamma$-ray emission from SFGs.} We show that the EGB is sensitive to the evolution of low mass populations of galaxies, particularly around $z\sim2.5$, and that it provides a new means to probe the engagement of CRs in these galaxies before the high noon of cosmic star-formation.
\end{abstract}


%

\section{Introduction}

The exact 
physical origin of the extra-galactic $\gamma$-ray background (EGB) remains unsettled. 
While a substantial fraction of the flux is attributed to blazars and radio galaxies~\citep[see, e.g.][]{Inoue2009ApJ, Singal2012ApJ, Ajello2015ApJ, Inoue2011ApJ, DiMauro2014ApJ, DiMauro2014JCAP, Wang2016NatPh, Stecker2019arXiv}, the detection of 
several nearby star-forming galaxies (SFGs) in $\gamma$-rays~\citep{Ajello2020ApJ_SFG} have also established these as a candidate EGB source class, where their $\gamma$-ray emission has been linked to their star-formation activity {through cosmic ray processes~\citep{Ha2021ApJ, Ambrosone2021ApJ}}. 
The EGB is comprised of two components: a contribution from resolved extragalactic sources, and an isotropic component from all other unresolved $\gamma$-ray emitting sources beyond our Galaxy, extending over redshift to the furthest reaches of the observable Universe.\footnote{This unresolved component is diffuse and appears isotropic on large-scales, sometimes being referred to as the isotropic $\gamma$-ray background (IGRB, e.g. as in~\citealt{Ackermann2015ApJ}). In this work, we put focus on the isotropic, diffuse $\gamma$-ray background component, and refer to this simply as the EGB, unless otherwise specified.}
While the origin of the resolved component can be readily decomposed into populations of individual sources (e.g. bright $\gamma$-ray emitters such as blazars; see~\citealt{Ajello2015ApJ}), the unresolved component could be comprised of a broader range of phenomena. This can include numerous blazars and radio galaxies, including those of low luminosity or at distances where they cannot be resolved as individual point sources~\citep[see, e.g.][]{Inoue2009ApJ, Singal2012ApJ, Ajello2015ApJ, Inoue2011ApJ, DiMauro2014ApJ, DiMauro2014JCAP, Wang2016NatPh, Stecker2019arXiv}, populations of active galactic nuclei (AGN) that do not exhibit relativistic jets~\citep{Tamborra2014JCAP, Wang2016NatPh}, {galaxy clusters~\citep[e.g.][]{Hussain2022arXiv}}, diffuse processes such as annihilating or decaying dark matter particles~\citep[see][]{Fornasa2015PhR}, or SFGs {(including high-energy emission from their galactic winds and associated circumgalactic structures -- see, e.g.~\citealt{Peretti2022MNRAS}).} 

Many earlier studies of the SFG contribution to the {isotropic} $\gamma$-ray background typically found that their contribution would be sub-dominant, between 10\% and 50\% of the total EGB intensity~\citep{Fields2010ApJ, Stecker2011ApJ, Makiya2011ApJ, Chakraborty2013ApJ, Tamborra2014JCAP}. Of these, some works~\citep[e.g.][]{Chakraborty2013ApJ, Makiya2011ApJ} further sought to differentiate between contributions from SFGs according to their mode of star-formation 
(main sequence or starburst, where main sequence star-formation is {relatively slow}, extended throughout the disk of a galaxy and arises over 1-2 Gyr timescales, while the starburst mode is more rapid, concentrated in the nuclear region of the galaxy and substantially more intense; see e.g. \citealt{Elbaz2011AA, Rodighiero2011ApJ, Schreiber2015AA}). These showed that the contributed flux from starburst galaxies was actually relatively minor compared to the total contribution from SFGs (main sequence and starburst), however conclusions 
could differ depending on exactly how the distinction between starburst and main sequence galaxies was drawn~\citep[e.g.][which found a comparatively larger contribution from starbursts, particularly at higher energies]{Sudoh2018PASJ}.

\begin{table*}
\centering
\begin{tabular}{*{4}{c}}
\midrule
Parameter & Value & Definition & Reference \\
\midrule
$\Gamma$ & $-2.1$ & CR proton spectral index & \cite{Ajello2020ApJ_SFG} \\
$\gamma_{\rm p}^{\star}$ & $10~\text{PeV}/m_{\rm p}c^2$ & Maximum CR proton energy & \cite{Peretti2019MNRAS} \\
$f_{\rm adv}$ & {0.2} & {Retained fraction of CR protons in a SFG after losses to advection} & \cite{Lacki2011ApJ} \\
$D_0$ & $3.0\times 10^{28}$ cm$^2$ s$^{-1}$ & CR diffusion coefficient normalisation & \cite{Aharonian2012SSRv} \\
$f_t$ & 0.1 & Efficiency of energy transfer from turbulent kinetic energy to magnetic energy & \cite{Federrath2011PRL} \\
$\alpha$ & {0.05} & Fraction of stars that produce a core-collapse SN event & \cite{Owen2018MNRAS} \\
$M_{\rm SN}$ & $50 \;\! {\rm M}_{\odot}$ & Upper mass of stars able to produce a SN event & \cite{Fryer1999ApJ, Heger2003ApJ} \\
$E_{\rm SN}$ & $10^{53}\;\!{\rm erg}$ & Total energy of a core-collapse SN & \cite{Smartt2009ARAA} \\
$\varepsilon$ & 0.1 & CR acceleration efficiency in SN remnants & \cite{Morlino2012AA}  \\
$f_{\nu}$ & $0.01$ & {SN kinetic energy available after losses to neutrinos} & \cite{Smartt2009ARAA} \\
{$\kappa_{\rm e}$} & {$0.034$} & {Fraction of total CR energy supplied to primary electrons} & \cite{Persic2014AA} \\
$f_{\rm abs}$ & 0.26 & Fraction of ionising stellar photons absorbed by interstellar Hydrogen & \cite{Petrosian1972ApJ} \\
$\beta$ & 0.6 & Average dust absorption efficiency of non-ionising photons & \cite{SavageARAA1979} \\ 
$\eta$ & 0.5 & Fraction of infra-red emission from diffuse interstellar gas & \cite{Helou1986ApJ} \\
$T^{\star}$ & $30,000\;\!{\rm K}$ & Temperature of the stellar radiation field & Same as~\citetalias{Owen2021MNRAS} \\
\midrule 
\end{tabular}
\caption{A list of fixed physical parameters adopted in our prototype galaxy model to specify its emitted $\gamma$-ray spectrum. Other model parameters are varied according to the physical properties of the galaxies.}
\label{tab:param}
\end{table*}

More recently, the SFG contribution to the {isotropic} EGB has been revisited. 
\cite{Peretti2020MNRAS} invoked a \textit{prototype} approach to model the SFG  contribution to the {diffuse} EGB spectrum, where a spectral model based on the nearby starburst galaxy M82 was scaled according to star-formation rate. It was assumed that SFGs would otherwise have the same physical properties as the M82 reference. An {isotropic} EGB spectrum was then modelled by convolution of this prototype with an appropriate star-formation rate function, and integrating over redshift. \citet{Ambrosone2021MNRAS} adopted a similar approach, but introduced some new refinements. In particular, they considered a blended range of spectral indices for the internal CR protons. By averaging over a distribution of indices informed by those observed in nearby SFGs~\cite[as listed in][]{Ajello2020ApJ_SFG}, they showed the resulting {isotropic} EGB intensity and spectrum could be modified compared to the fixed-spectrum prototype of \cite{Peretti2020MNRAS}.

A further model was later introduced by~\citep{Owen2021MNRAS}, hereafter~\citetalias{Owen2021MNRAS} (and subsequently extended in~\citealt{Owen2021PoS}), which included a treatment for CR abundance determined from the SFR of a galaxy, and self-consistently computed the SFG $\gamma$-ray emission spectrum (including internal attenuation effects from pair-production processes in interstellar radiation fields). However, certain parameters for their prototype model were assigned fixed, fiducial values - namely the size of the starburst core, gas density and CR escape fraction, and variations of their fiducial values were shown to have discernible impacts on the {diffuse} EGB spectrum. 
The approach adopted by \citet{Roth2021Natur} introduced a more refined physical model, including a galaxy-by-galaxy determination of an energy-dependent CR calorimetry fraction, allowing for a more physically-informed assessment of the $\gamma$-ray emission spectrum from individual galaxies. The number of fiducial parameters was reduced to just three inputs: the CR injection spectral index, the energy per supernova event, and the fraction of supernova energy that goes into primary CR ions and electrons. Other inputs to their model were informed by observations, and the distribution of SFGs was sampled from galaxy survey data. This showed a substantially higher {diffuse} EGB contribution from SFGs than earlier works, and could alone fully account for the isotropic $\gamma$-ray background. 

CR interactions lead to the deposition of momentum and energy in their host environment~\citep[e.g.][]{Owen2018MNRAS, Tibaldo2021Univ}. Thus, they can become important in 
controlling the evolution of galaxies. Using energetic backgrounds like the EGB to probe the engagement of CRs in their host galaxies over redshift is therefore valuable, and can provide crucial insights into the role of CRs in regulating the evolution of galaxies over cosmic time. 
\citetalias{Owen2021MNRAS} showed that characteristic separation of galaxies at a given redshift, described by the galaxy power spectrum~\citep{Tegmark2004PhRvD}, would imprint
a spatial signature into the EGB according to the redshift of the source population. The resulting angular power spectrum of the small-scale EGB anisotropies is thus sensitive to the properties and redshift distribution of the source population class. Analysis of these small-scale anisotropic patterns could offer insights into the evolution of CR engagement in SFG populations, and even allow different EGB source populations with different redshift distributions to be distinguished~\citep[e.g.][]{Ackermann2018PhRvL}.

In this work, we build on the earlier results of \citetalias{Owen2021MNRAS} to investigate the detailed imprints of SFGs in the EGB, and discern the characteristics of 
those SFGs contributing most strongly.
This may be considered a refinement of the approach adopted in ~\citetalias{Owen2021MNRAS}, where physical assumptions made in their fiducial prototype model are now relaxed so as to allow the $\gamma$-ray prototype emission spectrum to be more self-consistently determined from individual galaxy properties. This is intended to yield a more physical model, which can be used to resolve and investigate the contributions of different sub-classes of SFGs to the {isotropic} EGB, segregated according to their physical characteristics. 

We arrange this paper as follows. In section~\ref{sec:methods}, we describe our model for the $\gamma$-ray emission from a SFG, and outline the differences compared to the prototype model of~\citetalias{Owen2021MNRAS}. In section~\ref{sec:populations}, we introduce our models for galaxy populations, their physical properties and our criteria for 
{distinguishing starbursts from the broader population source population of SFGs.} Our results are presented in section~\ref{sec:results} with discussion, and we draw conclusions in section~\ref{sec:conclusions}. 

\section{Methodology}
\label{sec:methods}

We construct a model for the EGB contribution from SFGs using a prototype model based on that introduced in \citetalias{Owen2021MNRAS}. We refer the reader to this earlier work for a detailed description of our code, galaxy model, CR interaction and $\gamma$-ray emission model and numerical techniques. Here we provide an overview of the model, and emphasise the aspects that are different from the previous work. A summary of the fixed model parameters in shown in Table~\ref{tab:param}, with their adopted fiducial values. Note that some model parameters (e.g. SFG size, $R$, galaxy stellar mass $M^{\star}$, and mean interstellar gas density $\langle n_{\rm H} \rangle$) which were previously fixed to fiducial values in \citetalias{Owen2021MNRAS} are now determined self-consistently from physical galaxy properties provided by our galaxy population models (see section~\ref{sec:populations}). In the following, we  express particle energies in terms of their Lorentz factor, e.g. for protons, the total energy $E_{\rm p} = \gamma_{\rm p} \;\!m_{\rm p}c^2$. Photon energies (including $\gamma$-rays) are expressed as dimensionless quantities normalised to the electron rest mass energy, i.e. $\epsilon_{\gamma} = E_{\gamma}/m_{\rm e}c^2$, unless otherwise specified.

\subsection{Prototype galaxy model}

\subsubsection{Cosmic ray spectrum and energy budget}

High-energy $\gamma$-ray emission from SFGs is primarily driven by hadronic CR interactions. These can proceed through various channels, 
however
the internal conditions of typical star-forming galaxies would favour proton-proton (hereafter pp) pion-production processes~\citep{Owen2018MNRAS, Owen2019AA}. {An additional leptonic component arising from bremsstrahlung and inverse Compton scattering of electrons within their host galaxy would also be present~\citep[e.g.][]{Chakraborty2013ApJ, Pfrommer2017ApJ, Roth2021Natur}, and this can be enhanced by the secondary CR electrons supplied by pp interactions.}  
The pp interaction arises above a threshold proton kinetic energy of $\sim 0.28~\text{GeV}/c^2$, and leads to the formation of charged and neutral pions. The neutral pions decay to form $\gamma$-rays. The interaction rate is given by 
  $\dot{n}_{\rm p\pi}(\gamma_{\rm p}) = \langle n_{\rm H}\rangle \;\! n_{\rm p}(\gamma_{\rm p}) \;\!{c}\;\!\sigma_{\rm p\pi}(\gamma_{\rm p})$, 
where $n_{\rm p}$ is the CR proton density, $\sigma_{\rm p\pi}$ is the total inelastic pp interaction cross section, 
and 
$\langle n_{\rm H}\rangle$ is the average ambient gas density within the host galaxy. 
In the present work, 
we model the gas density within the host galaxy 
using the physical galaxy properties. In particular, we estimate the molecular gas density, which we consider would form the primary target for hadronic CR interactions
(see section~\ref{sec:gas_density} for details). This is different to \citetalias{Owen2021MNRAS}, where a mean density was adopted with a single fiducial value of $1~{\rm cm}^{-3}$ applied uniformly to all SFG sources. 

The CR proton density model we use here is the similar to the prescription adopted in the earlier work, where the steady-state proton spectrum in a SFG followed from the model introduced in~\cite{Owen2019AA}. This balanced the injection rate of CRs (assumed to scale with the SN event rate of a galaxy) with their absorption by hadronic interactions and diffusive and advective escape~\citep[see also][]{Owen2021PoS}, i.e.
\begin{equation}
       n_{\rm p}(\gamma_{\rm p})\;\!{\rm d}\gamma_{\rm p} = \; \frac{35 R^2 \;\! f_{\rm adv}\;\!\mathcal{L}_{0} \;\! \mathcal{A}{(\gamma_{\rm p})}}{108 \;\! D(\gamma_{\rm p})} \;\! \frac{\partial}{\partial \gamma_{\rm p}} \left(\frac{\gamma_{\rm p}}{\gamma_{\rm p,0}}\right)^{-\Gamma} \;\!{\rm d}\gamma_{\rm p} \ ,
       \label{eq:cr_spec1}
   \end{equation}
  where $\gamma_{\rm p, 0} = E_0/m_{\rm p} c^2 = 1 \;\! {\rm GeV}$ is used as a reference CR energy. 
  Here, $R$ is the size of the SFG. Previously, this was set to a fiducial value (of 0.1 kpc) for all galaxies. However, in this work we 
  model this using an effective galaxy size based on the star-formation rate, stellar mass and redshift of each galaxy (see section~\ref{sec:galaxy_size} for details). 
  $f_{\rm adv}$ is the fraction of CRs {retained within a galaxy after accounting for losses} by advection (presumably in a galactic outflow). {These advected CRs} do not engage in hadronic interactions within the SFG, and therefore do not contribute to the $\gamma$-ray emission {in our model}. 
  {Dynamical flows in and around SFGs has been demonstrated to 
  modify their $\gamma$-ray emission~\citep[see, e.g.][]{Kornecki2022AA, Peretti2019MNRAS}. Detailed investigation of the impacts of these effects on the EGB is worthy of dedicated study, but falls beyond the focus of our present work. As such, we adopt a fiducial value for $f_{\rm adv} = 0.2$ to approximate their effects. This choice reflects the lower end of the range of calorimetric fractions suggested by~\cite{Lacki2011ApJ}, and
  and is lower than the value of 0.5 adopted previously  in~\citetalias{Owen2021MNRAS}. We consider our more conservative choice here to be more appropriate for our revised model, where $\gamma$-ray emission is dominated by relatively small galaxies with shallow potential wells and concentrated regions of strong star-formation (see section~\ref{sec:SFG_mass_contributions}). Such conditions would favour the development of strong, faster advective outflows~\citep[see, e.g.][]{Yu2020MNRAS} that would be more effective in removing CRs from their host galaxy.}
  
  We quantify the impact of CR escape losses
  due to diffusion using the parameter  
  $D(\gamma_{\rm p}) = D_0 ( {r_L(\gamma_{\rm p},\langle |B| \rangle)|}/{r_{L,0}})^{\varsigma}$, where $\langle |B| \rangle$ is the characteristic interstellar magnetic field strength in a SFG, which we approximate for each galaxy assuming a turbulent dynamo mechanism~\citep[see][]{Schober2013A&A}, where $\langle |B| \rangle = \left({4 \pi \; {\mu_{\rm H}}\;\!\langle n_{\rm H}\rangle }\right)^{1/2} v_{\rm f} f_t$. {Here, $\mu_{\rm H}=1.4\ m_{\rm p}$ is the mean molecular mass of interstellar gas, $m_{\rm p}$ is the proton rest mass,} 
  $f_t = 0.1$ \citep{Federrath2011PRL} is the efficiency of energy transfer from turbulent kinetic energy to magnetic energy, 
  and $v_{\rm f}$ as the fluctuation velocity of the galaxy under pressure-gravity equilibrium ($v_{\rm f}  \approx R ({2}\pi \rho G/3)^{1/2}$). 
  $\varsigma = 1/2$~\citep[e.g.][]{Berezinskii1990book} encodes the effect of interstellar magnetic turbulence on CR diffusion, {with our chosen value being} appropriate for a Kraichnan-type turbulence spectrum. The reference value of the coefficient, $D_0 = 3.0\times 10^{28}$ cm$^2$ s$^{-1}$ is specified for a 1 GeV proton diffusing through a 5$\mu$G magnetic field (with $r_{L, 0}$ as the corresponding gyro-radius of the particle). This is based on empirical estimates for the CR diffusion coefficient for the Milky Way~\citep[e.g.][]{Aharonian2012SSRv}. While the model prescription for CR advection and diffusion is the same as that used in~\citetalias{Owen2021MNRAS}, the dependency on quantities that are physically determined from the galaxy population model rather than universally-applied fiducial quantities will lead to some variation in the exact CR steady-state spectral normalisation in the current work. 
  
  The CR energy spectrum is modelled as a single power-law, of index $\Gamma$, which we set $\Gamma = -2.1$. This is characteristic of the mean of the range of values inferred for local SFGs detected in $\gamma$-rays~\citep{Ajello2020ApJ_SFG}. 
   The mean attenuation of protons $\mathcal{A}$ follows the \citetalias{Owen2021MNRAS} approximation, which is specified by the size of the galaxy $R$, its mean gas density $\langle n_{\rm H}\rangle$, and the energy-dependent total pp interaction cross section, for which we adopt the parametrization of~\citet{Kafexhiu2014}. As the galaxy size and mean density in this work are based on the galaxy properties, the hadronic attenuation factor in this work is more self-consistently estimated compared to the approach used in~\citetalias{Owen2021MNRAS}. 
   
   {$\mathcal{L}_{0}$ is the volumetric CR injection power, and is} set by the CR luminosity of a SFG. {This} is specified by its supernova (SN)
   event rate 
   and the CR energy injected by each SN event. We normalise this term to the volume of the host galaxy, as the distribution of CR injection within each SFG is not consequential to our calculations.  
   The SN event rate is related to the star-formation rate of a galaxy by $\mathcal{R}_{\rm SN} = \alpha \mathcal{R}_{\rm SF}/M_{\rm SN}$, where we set $\alpha = 0.05$~\citep[following][]{Owen2018MNRAS} for the fraction of stars which evolve to produce a core-collapse SN (which are expected to dominate SN activity in these highly star-forming systems), and $M_{\rm SN} = 50 M_{\odot}$ is the upper cut-off mass for stars able to produce a SN event~\citep{Fryer1999ApJ, Heger2003ApJ}. 
   The total energy of core-collapse SNe  is $E_{\rm SN} = 10^{53}\;\!{\rm erg}$~\citep[e.g.][]{Smartt2009ARAA}, and the efficiency of energy transfer from the SN to CRs is given by the product of $\varepsilon = 0.1$ (the CR acceleration efficiency -- see, e.g.~\citealt{Morlino2012AA}), {and
   the retained fraction of SN energy after losses to neutrinos $f_{\nu} = 0.01$}~\citep[see, e.g.][for a value appropriate for core-collapse SNe]{Smartt2009ARAA}.
   
\subsubsection{{Hadronic} $\gamma$-ray production in SFGs}
\label{sec:gamma_ray_from_gal}

{The hadronic $\gamma$-ray 
emission from a SFG arises
through the production and subsequent decay of neutral pions.} The $\gamma$-ray spectral emissivity by this channel is given by
\begin{equation}
\frac{{\rm d}\dot{n}_{\gamma}(\epsilon_{\gamma})}{{\rm d}\epsilon_{\gamma}} = 
c\;\! \langle n_{\rm H}\rangle \;\! \int_{\gamma_{\rm p}^{\rm th}}^{\gamma_{\rm p}^{\star}} \frac{{\rm d}{\sigma}_{\rm p\gamma}(\gamma_{\rm p}, \epsilon_{\gamma})}{{\rm d}\epsilon_{\gamma}}\;\! n_{\rm p}(\gamma_{\rm p}) \;\! {\rm d}\gamma_{\rm p} \ ,
\label{eq:emiss_spec}
\end{equation}
(see~\citetalias{Owen2021MNRAS}), where $n_{\rm p}$ is the CR density, given by equation~\ref{eq:cr_spec1},  $c$ is the speed of light, and an upper CR proton energy limit is set as $\gamma_{\rm p}^{\star} = 10~\text{PeV}/m_{\rm p}c^2$~\citep[][]{Peretti2019MNRAS}. ${{\rm d}{\sigma}_{\rm p\gamma}(\gamma_{\rm p}, \epsilon_{\gamma})}/{{\rm d}\epsilon_{\gamma}}$
is 
the differential $\gamma$-ray inclusive cross section, for which we adopt the parametrization of \citet{Kafexhiu2014}. 

\subsubsection{{Leptonic $\gamma$-ray production in SFGs}}

{Energetic electrons can also drive $\gamma$-ray emission from SFGs. This primarily arises from inverse Compton scattering in ambient thermal radiation fields. Non-thermal bremsstrahlung emission can also make a non-negligible contribution. Together, the emission from these processes can form an important component of the sub-GeV $\gamma$-ray emission from SFGs~\citep[e.g.][]{Roth2021Natur}. This was not previously considered in the~\citetalias{Owen2021MNRAS} model, which put focus on the SFG emission at slightly higher energies, where hadronic emission would be more likely to dominate.}

CR electrons are supplied both by their direct acceleration (primary electrons), and also by the decay of charged pions in hadronic interactions (secondary electrons and positrons - both referred to hereafter as secondary electrons). The acceleration sites for primary electrons would presumably be the same as for CR protons. As such, {we relate their CR volumetric injection power $\mathcal{L}_{\rm e}$ to that of protons $\mathcal{L}_0$ by a factor 
of $\kappa_{\rm e} = \left(m_{\rm p}/m_{\rm e}\right)^{-(3+\Gamma)/2} \approx 0.034$, 
which follows from the ratio of energy passed to electron and proton energy densities at their acceleration site, obtained} by~\citealt{Persic2014AA} (see also~\citealt{Persic2015mgm}). {For the injection of secondary electrons from hadronic interactions, we compute the production fraction of electrons in a primary CR proton's rest frame via the pp interaction using the publicly available code {\tt aafragpy}~\citep{Koldobskiy2021}.}\footnote{{This code is based on {\tt Aafrag}~\citep{Kachelriess2019CoPhC}, but provides an extension to lower CR proton energies, below 4 GeV, using production parameterizations obtained by~\cite{Kamae2006ApJ, Kamae2007ApJ}.}}

{CR electrons cool more rapidly than CR protons, and practically would lose their energy within a SFG, even in the presence of a galactic outflow. As such, the escape fraction of CR electrons from a SFG is negligible and the steady-state mean density of CR electrons throughout a SFG is obtained from the balance between their injection rate and cooling timescale:
\begin{equation}
n_{\rm e}(\gamma_{\rm e}) {\rm d}\gamma_{\rm e} = \mathcal{Q}_{\rm e}(\gamma_{\rm e}) \;\! \tau_{\rm cool}(\gamma_{\rm e}) \;\! {\rm d}\gamma_{\rm e} \ ,
\end{equation}
where the normalization of the injection rate $\mathcal{Q}_{\rm e}$ is set by $\mathcal{L}_{\rm e}$ for primary electrons (assuming the same injection spectrum as adopted for CR protons), or the hadronic interaction rate and inclusive production spectra for secondaries~\citep{Berrington2003ApJ}. $\tau_{\rm cool}$ is the total energy-dependent cooling timescale, which accounts for CR energy losses to radiative (inverse Compton and synchrotron) cooling, bremsstrahlung, and Coulomb interactions.}

{Using the steady-state electron spectrum (including primary and secondary leptons), we computed the inverse Compton emission from a SFG following~\citet{Blumenthal1970RvMP}. We considered this to be dominated by up-scattered photons from the cosmological microwave background (CMB), with contributions from other interstellar thermal radiation fields (e.g. starlight) being relatively unimportant to the production of $\gamma$-rays over the energy range of interest ($\sim$0.01-50 GeV). To compute the bremsstrahlung emission spectrum, we followed the treatment of~\cite{Schlickeiser2002_book}, with the emission scaling with the SFG mean gas density. The total leptonic $\gamma$-ray emission then followed as the sum of the inverse Compton and bremsstrahlung emission spectra.}

\subsubsection{{$\gamma$-ray attenuation in SFG environments}}

The $\gamma$-rays {produced in a SFG} can be attenuated by $\gamma\gamma$ pair-production interactions in low energy radiation fields associated with the CMB, starlight, or reprocessed starlight by dust. The impact of this is small (although not entirely negligible) at $\gamma$-ray energies below $\sim 10$ GeV, but can severely attenuate the $\gamma$-ray emission from a SFG at TeV energies. 
Our treatment of this internal $\gamma$-ray attenuation process is identical to that of~\citetalias{Owen2021MNRAS}, however the geometric dilution of stellar radiation fields is modelled as a diluted black-body spectrum, and computed for each galaxy according to its derived size, rather than using a fiducial radiation field volume as in the previous work. As before, the interstellar dust temperature is specified by the galaxy redshift, according to the empirical relation of~\citet{Schreiber2018AA}, the infra-red dust luminosity is scaled by star-formation rate according to~\citet{Kennicutt1998ApJ}, and this is also used to specify the total stellar radiative output power of stars in a SFG using the relation of \cite{Inoue2000PASJ}, with a fraction $f_{\rm abs} = 0.26$ of ionising stellar photons absorbed by interstellar  Hydrogen~\citep{Petrosian1972ApJ}, an average dust-absorption efficiency of non-ionising photons from central sources in ionised, star-forming regions of $\beta = 0.6$~\citep{SavageARAA1979}, and a fraction of 
$\eta = 0.5$ of the infra-red emission being attributed to diffuse interstellar gas, rather than from star-forming regions~\citep{Helou1986ApJ}.
We set the temperature of the stellar radiation field to be $T^{\star} = 3\times10^4~{\rm K}$, to reflect the temperature of a dominant stellar population of O/B-type stars typical of SFGs. 

\subsection{Cosmological $\gamma$-ray propagation}
\label{sec:cosmo_rt}

We model the propagation of $\gamma$-rays from populations of SFGs to form an EGB model at $z=0$ using a cosmological radiative transfer approach, which ensures conservation of photon number and phase space volume~\citep{Fuerst2004AA, Chan2019MNRAS}. This accounts for pair-production processes arising between $\gamma$-rays and soft, intergalactic extra-galactic background light (EBL) photons, and the subsequent inverse Compton scattering of CMB photons to $\gamma$-ray energies by the produced pairs (the \textit{cascade} effect), using the semi-analytic EBL model of~\cite{Inoue2013ApJ}. While this effect does not have a large impact on our results, minor influences can be seen at the highest energies we consider in some of our spectra, where some attenuation is evident.  We integrate our model over a redshift range between $z_{\rm max} = 3$ and $z_{\rm obs} = 0$, assuming a flat Friedmann-Robertson-Walker cosmology with cosmological parameters from~\cite{Planck2018}. The computational implementation of this method is identical to that used in~\citetalias{Owen2021MNRAS}.

\subsection{EGB anisotropies and spectrum}
\label{sec:anisotropies_egb}

Although individual sources would not be resolved, small-scale anisotropic signatures are imprinted into the EGB by the spatial distribution of SFGs.\footnote{For analyses 
of such anisotropies in 
\textit{Fermi}-LAT EGB data, see~\citet{Ackermann2012PhRvD, PeerboomsPoS2021}. Anisotropies were also used to identify two EGB source classes in~\cite{Fornasa2016PhRvD} and~\cite{Ackermann2018PhRvL}. Additionally, prospects for EGB anisotropy studies with future facilities are discussed in~\citet{Hutten2018JCAP}.}
The clustering of galaxies is a biased tracer of the underlying dark matter distribution of the Universe. An effective clustering bias factor of SFGs compared to dark matter can be defined using the relation $P_{\rm g}(k, z) = b_{\rm SFG}(z)\;\!P_{\rm lin}(k, z)$, where 
$P_{\rm g}(k, z)$ is the power spectrum of SFGs,
and $P_{\rm lin}(k, z)$ is the power spectrum of linear dark matter density fluctuations. We adopt the approximation of~\cite{Eisenstein1999ApJ} for $P_{\rm lin}(k, z)$, while the SFG population bias factor, $b_{\rm SFG}$ follows from the best-fit values of~\cite{Hale2018MNRAS}. The redshift dependence in the SFG power spectrum would imprint $\gamma$-ray intensity into the EGB at a corresponding scale. The strength of the contribution from SFGs located at a particular redshift would be set by the $\gamma$-ray luminosity of the source population at that epoch, and this would be discernible in the EGB by the spatial scale of that imprint. 
This could be measured in EGB observations using an auto-correlation function of the $\gamma$-ray sky intensity distribution. From this, 
clustering $\mathcal{C}_{\ell}^{C}$ and 
isotropic Poisson noise terms $\mathcal{C}_{\ell}^{P}$
 can be extracted using a Fourier Transform. Following~\citetalias{Owen2021MNRAS}, these may be written directly, as:
\begin{equation}
\mathcal{C}_{\ell}^{C}(E_{\gamma}) =  \int_0^{z_{\rm max}} \frac{{\rm d}^2V_{\rm c}}{{\rm d}z\;\!{\rm d}\Omega}{\rm d}z\;\ P\left(\frac{\ell_p}{r_p}[1+z]\right)\left\{ \frac{{\rm d}F_{\gamma}(E_{\gamma}, z)}{{\rm d}E_{\gamma}} \right\}^2 \ , 
\label{eq:differential_anisotropy}
\end{equation}
and
\begin{equation}
\mathcal{C}_{\ell}^{P}(E_{\gamma}) =  \int_0^{z_{\rm max}} \frac{{\rm d}^2V_{\rm c}}{{\rm d}z\;\!{\rm d}\Omega}{\rm d}z\;\! \left\{ \frac{{\rm d}F_{\gamma}(E_{\gamma}, z)}{{\rm d}E_{\gamma}} \right\}^2 \ ,
\label{eq:final_poisson}
\end{equation}
respectively, in differential units of flux, where ${\rm d}F_{\gamma}/{{\rm d}E_{\gamma}}$ is the redshift-dependent $\gamma$-ray flux from a population of SFGs, computed using the cosmological radiative transfer approach described in section~\ref{sec:cosmo_rt}.
 In order to model observationally practical signatures, we later integrate the anisotropic signatures over energy bands to reduce the requirement on photon numbers in small ranges of photon energy. 
 The spectrum of the EGB is also modelled. This follows simply as the differential $\gamma$-ray flux contribution integrated over the set redshift range. Both the EGB anisotropy and spectrum calculation method are the same as those used in~\citetalias{Owen2021MNRAS}.
 
\section{Galaxy population model}
\label{sec:populations}

\subsection{Physical properties}
\label{sec:physical_properties_populations}

While~\citetalias{Owen2021MNRAS} investigated the star-formation rate distribution of galaxies over redshift to 
characterise their 
$\gamma$-ray contribution to the EGB, other physical properties of galaxy populations were fixed at fiducial values. {Although} this was sufficient for a first model to broadly characterize the nature of imprinted EGB signatures from SFGs, 
the complex inter-dependencies 
between star-formation rates, redshift evolution and certain physical properties of galaxies that would affect their overall $\gamma$-ray luminosity was left unexplored. 

In this work, we relax some of the fixed fiducial parameter values (and corresponding assumptions) used previously, and introduce a refined model to investigate the more detailed signatures that would arise from the redshift evolution of galaxy physical properties. This also opens-up the possibility to investigate the relative contributions of different classes of SFGs, and the evolving $\gamma$-ray emission from SFGs over cosmic time. \citetalias{Owen2021MNRAS} treated all SFGs equally in terms of their size and mass. Thus, the differences in the $\gamma$-ray contribution from higher mass compact galaxies and/or starbursts was left unresolved from the contributions made by lower mass and/or main sequence systems with similar star-formation rates, despite differences in their interstellar medium density and internal star-formation distribution that would likely impact very substantially on their $\gamma$-ray luminosity~\citep[see, e.g.][]{Sudoh2018PASJ, Roth2021Natur}.  
The following sub-sections outline the physical characteristics of galaxy populations included in our model. 

\subsubsection{Star-formation rate}

We model the star-formation rate function (SFRF) of galaxies using the reference model \verb|100N1504-Ref| of~\cite{Katsianis2017MNRAS}. This is the same approach as adopted in~\citetalias{Owen2021MNRAS}, and gives the number density of galaxies per decade in star-formation rate $\mathcal{R}_{\rm SF}$  obtained from simulations using the Virgo Consortium's Evolution and Assembly of GaLaxies and their Environments (EAGLE) project~\citep{Schaye2015MNRAS, Crain2015MNRAS}. This provides a broad coverage of SFRs, from {nearly dead, quenched} galaxies of $\mathcal{R}_{\rm SF} \sim 10^{-3} \;\!{\rm M}_{\odot}\;\!{\rm yr}^{-1}$ to very active starbursts of $\mathcal{R}_{\rm SF} \sim 10^{3} \;\!{\rm M}_{\odot}\;\!{\rm yr}^{-1}$, and extends up to $z\sim 8$ thus covering the range of interest in this work. Note that we re-scale the adopted SFRF to ensure consistent 
assumption of a \citet{Salpeter1955ApJ} stellar initial mass function (IMF) throughout all components of our model. 

\subsubsection{Stellar mass}
\label{sec:galaxy_stellar_mass_function}

We model the evolving galaxy stellar mass function (GSMF) following the best-fit double Schechter function presented by~\citet{McLeod2021MNRAS}. This is derived from a combination of \textit{Hubble Space Telescope} imaging surveys, and has sufficient redshift coverage for our model, reaching up to $z\sim 3.75$, although it is noted that higher redshift constraints on the GMSF suggest this parametrization remains reasonable up to $z\sim 5$~\citep{Duncan2014MNRAS, Grazian2015AA, Song2016ApJ}. While~\citet{McLeod2021MNRAS} considered separate fits to both SFGs and passive galaxies, we adopt their total GSMF best-fit model. {This retains all the information included in their separated `\textit{star-forming}' and `\textit{quiescent}' fits (if following the terminology of~\citealt{McLeod2021MNRAS}), thus allowing us to later set our own physically-motivated separation criteria for starburst and main sequence SFGs (see section~\ref{sec:sb_and_ms})}, rather than the observationally-motivated $UVJ$ colour-colour criteria~\citep[see also][]{Williams2009ApJ, Carnall2018MNRAS, Carnall2020MNRAS} used to separate the data sample in~\citet{McLeod2021MNRAS}. The GSMF double~\citet{Schechter1976ApJ} function model is defined as 
\begin{align}
    \phi(\mathcal{M}) = \ln(10) \cdot & \exp \left[-10^{(\mathcal{M}-\mathcal{M}^{\star})}\right] \cdot 10^{(\mathcal{M}-\mathcal{M}^{\star})} \nonumber \\
    & \cdot \left[\phi_1^{\star}\cdot 
    10^{(\mathcal{M}-\mathcal{M}^{\star})\alpha_1} + 
    \phi_2^{\star}\cdot 10^{(\mathcal{M}-\mathcal{M}^{\star})\alpha_2}\right] \ ,
\end{align}
which gives the number density of galaxies per dex in stellar mass, where $\mathcal{M} = \log_{10}(M^{\star}/{\rm M}_{\odot})$. The two components of the double Schechter function have the same characteristic stellar mass, $\mathcal{M}^{\star}$, which is redshift-dependent. The best-fit functional forms for $\mathcal{M}^{\star}$, $\alpha_1$ (the high-mass slope), $\alpha_2$ (the low-mass slope)
and the high and low mass normalisations ($\phi_1^{\star}$ and $\phi_2^{\star}$, respectively) are given by:
\begin{align}
    \mathcal{M}^{\star} & = a_1 + a_2 z \ ,  \label{eq:1star}\\
    \alpha_1 & = a_3 + a_4 z \ , \\
    \alpha_2 & = a_5 + a_6 z \ , \\
    \log (\phi_1^{\star}) & = a_7 + a_8 z + a_9 z^2 \ , \\
    \log (\phi_2^{\star}) & = a_{10} + a_{11} z \ , 
    \label{eq:4star}
\end{align}
where the best-fit parameters $a_1$ to $a_{11}$ are provided in Table~\ref{tab:fitting_params} for intrinsic galaxy parameter values. 

\begin{table}
\centering
\begin{tabular}{*{2}{c}}
\midrule
{Fitting parameter} & {Value} \\
\midrule
$a_1$ & 10.55 \\
$a_2$ & 0.0 \\
$a_3$ & -0.16 \\
$a_4$ & 0.12 \\
$a_5$ & -1.45 \\
$a_6$ & -0.08 \\
$a_7$ & -2.43 \\
$a_8$ & -0.17 \\
$a_9$ & -0.08 \\
$a_{10}$ & -2.94 \\
$a_{11}$ & -0.22 \\
\midrule 
\end{tabular}
\caption{Best-fit intrinsic galaxy parameter values for equations~\ref{eq:1star} to~\ref{eq:4star}, following the total evolving GSMF parameterization of~\citet{McLeod2021MNRAS}.}
\label{tab:fitting_params}
\end{table} 

\subsubsection{Effective size}
\label{sec:galaxy_size}

The definition of the size of a galaxy is ambiguous. However, for the purposes of this work, we require an effective galaxy size that is physically useful to estimate the mean gas density of the interstellar medium, the CR number density (cf. equation~\ref{eq:cr_spec1}, which is indirectly dependent on the galaxy effective size $R$), and the distance over which the $\gamma\gamma$ internal attenuation can be estimated. We consider that stellar light-weighted sizes are sufficient for this estimate, as they characterise (1) the distribution of stars within a SFG over which stellar end-products are distributed (and thus the distribution of CR injection throughout the galaxy), (2) the distribution of star-light and, presumably, dust-reprocessed starlight throughout the galaxy, which is the physical quantity required to estimate $\gamma$-ray internal absorption effects, and (3) the distribution of gas where the stars and CRs are located, which is most relevant for the production of $\gamma$-rays in a SFG. \cite{Wel2014ApJ} showed that galaxies have a tight scaling relation with galaxy stellar mass, but {that} SFG and {quenched galaxy} population scaling relations are distinct from one another (for an overview, see also~\citealt{ForsterSchreiber2020ARA&}). In both cases, the scaling relation may be written in the form 
\begin{equation}
R(M^{\star}, z) = R_e(z) \left( \frac{M^{\star}}{M^{\star}_{\rm 0}}\right)^{\beta} \ ,
\label{eq:size_mass_relation}
\end{equation}
where we use a reference mass of ${M^{\star}_{\rm 0}} = 5\times 10^{10}\;\!{\rm M}_{\odot}$, and $\beta = 0.22$ for SFGs, or $\beta = 0.75$ for {quenched} galaxies.\footnote{Note that this separation between SFGs and {quenched} galaxies is in addition to the sub-division of the SFGs into starbursts and main sequence SFGs described in section~\ref{sec:sb_and_ms}.} The redshift-dependence is given by
\begin{equation}
    R_e(z) = R_0 (1+z)^{-\delta} \ .
\end{equation}
For SFGs, \cite{Wel2014ApJ} found $\delta = 0.75$, and that $R_0 = 7.24$ kpc at $z=0.25$. Conversely, for {quenched} galaxies, they found $\delta = 1.48$ and $R_0 = 3.98$ kpc. For completeness, we consider both the SFGs and {quenched} galaxies in our calculations, distinguishing between the two populations by their specific star-formation rate, ${\rm sSFR} = \mathcal{R}_{\rm SF}/M^{\star}$~\citep[we note that, in contrast][separated SFGs from {quenched} galaxies using a $U-V$ vs. $V-J$ colour-colour method]{Wel2014ApJ}, where SFGs have ${\rm sSFR}\geq 10^{-11}\;\!{\rm yr}^{-1}$~\citep[e.g.][]{Merlin2018MNRAS}. 
However, the $\gamma$-ray emission from {quenched} galaxies is very low compared to their SFG counterparts, and we find their contribution to the EGB to be inconsequential. 

\subsubsection{Molecular gas density}
\label{sec:gas_density}

As neutral gas dominates the mass of the interstellar medium of a galaxy, it provides the main target for the hadronic interactions that drive $\gamma$-ray production~\citep{Roth2021Natur}. 
This is typically concentrated in molecular clouds and dense structures within the interstellar medium of a galaxy, with such structures in the Milky Way having been identified as CR interaction targets and, hence, $\gamma$-ray sources at GeV to TeV energies (\citealt{Gabici2007ApSS, Ackermann2012ApJb, Ackermann2012ApJc,Yang2014AA, Tibaldo2015ApJ, Dogiel2018ApJ}; see~\citealt{Tibaldo2021Univ} for a review), even offering potential as 
an probe of particle acceleration in their vicinity~\citep{Mitchell2021MNRAS}. Thus, much of the $\gamma$-ray emission from a SFG could be expected to arise from CR interactions within molecular clouds~\citep{Peng2019AA}. The exact determination between the $\gamma$-ray emission and star-formation rate would depend on the ability of the CRs to propagate into these dense molecular regions, and this can vary between individual clouds due to their structure and magnetic field configuration~\citep[e.g.][]{Dogiel2018ApJ, Owen2021ApJ}. However, scaling models between tracers of molecular gas and $\gamma$-ray emission have seen sufficient success in recent works~\citep[e.g.][]{Rojas2016MNRAS, Ajello2020ApJ_SFG, Peng2019AA} to justify leaving these more detailed cloud-scale propagation considerations to future studies.

We consider that the $\gamma$-ray emission region of a SFG is dominated by its molecular gas component.
We thus calculate the 
mean density parameter for a galaxy $\langle n_{\rm H} \rangle$ (as used in equation~\ref{eq:emiss_spec}) using 
estimates for its molecular gas content. 
This is closely connected to the star-forming activity of a galaxy, $\mathcal{R}_{\rm SF}$, which is typically fuelled by a rich supply of molecular gas, and suitable empirically-obtained scaling relations between $\mathcal{R}_{\rm SF}$ and estimated molecular gas density are widely available. 
We adopt the star-formation law  of~\cite{Leroy2013AJ}, which is 
empirically obtained from the observed 
star-formation surface density $\Sigma_{\rm SFR} = \mathcal{R}_{\rm SF}/\pi R^2$ and molecular gas surface density $\Sigma_{\rm H_2} = M_{\rm H_2}/\pi R^2$ (here, $M_{\rm H_2}$ is the total molecular gas mass of the galaxy) in nearby spiral galaxies. This gives:
\begin{equation}
    \Sigma_{\rm H_2} = 10 \;\! {\rm M}_{\odot} {\rm pc}^2\;\! \left(\frac{\Sigma_{\rm SFR}}{q \;\! {\rm M}_{\odot} {\rm yr}^{-1} {\rm kpc}^{-2}}\right)^{1/\upsilon} \ ,
    \label{eq:surface_density}
\end{equation}
where the values of $q$ and $\upsilon$ were 
determined by Monte Carlo fitting to observations with different combinations of star-formation and molecular gas tracers~\citep{Leroy2013AJ}. We convert this to an estimate for the mean gas density of a SFG by 
$\langle n_{\rm H} \rangle \approx 3 \Sigma_{\rm H_2}/4 R m_{\rm H}$, where $m_{\rm H}$ is the atomic mass of Hydrogen. 

In equation~\ref{eq:surface_density}, values of 
$q = 4.47\times 10^{-3}$ and $\upsilon = 0.90$ follow from the combined use of CO and H$\alpha+24 \;\! \mu$m as a molecular gas tracer~\citep{Leroy2013AJ}, where
a CO to H$_2$ conversion factor assuming a minimum surface density of molecular clouds of $50\;\! {\rm M}_{\odot} {\rm pc}^{-2}$ is used. Other studies have considered a higher floor surface density for this conversion factor, of~$100\;\! {\rm M}_{\odot} {\rm pc}^{-2}$~\citep[e.g.][]{Narayanan2012MNRAS}, which has been considered as a more appropriate limit for star-forming galaxies~\citep{Hughes2013ApJ}. While a fit for a $100\;\! {\rm M}_{\odot} {\rm pc}^{-2}$ cut-off is also available in~\citep{Leroy2013AJ}, we consider the lower choice more appropriate for this study, 
as it 
would seem to be more sensitive to lower density `translucent' gas clouds~\citep[e.g.][]{Heyer2009ApJ, Liszt2010AA}, which would presumably also comprise a substantial part of the $\gamma$-ray emission volume of a galaxy. 

\subsection{Starburst and main sequence SFGs}
\label{sec:sb_and_ms}

\begin{figure}
    \centering
    \includegraphics[width=\columnwidth]{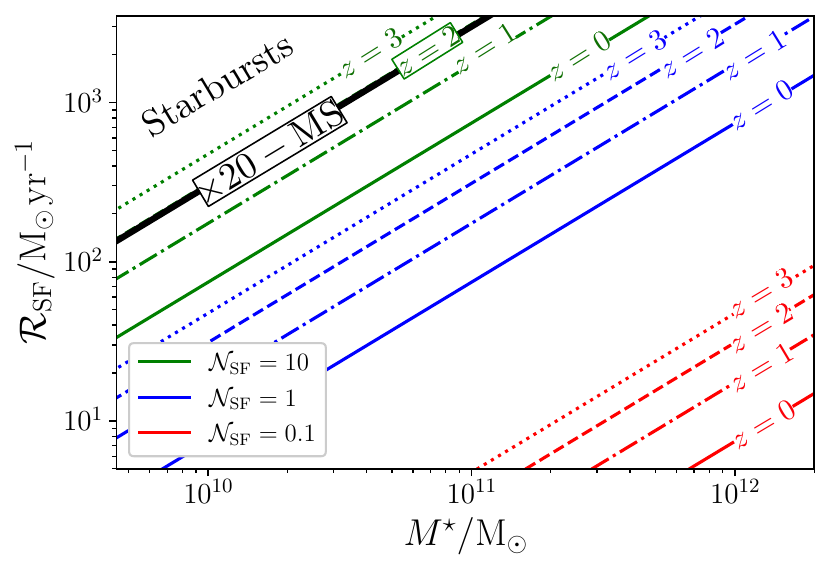}
    \caption{{Illustration of different choices of the starburst/main sequence separation criteria $\mathcal{N}_{\rm SF}$ on the 
    $\mathcal{R}_{\rm SF}-M^{\star}$ plane at different redshifts (as labeled). Our choice of $\mathcal{N}_{\rm SF}\geq 10$ for starburst SFGs is shown by the green lines, and is broadly consistent with a level that is 20$\times$ the 
    {star-formation rate of a main sequence star-forming galaxy at fixed stellar mass, for the main sequence} determined for galaxies at $z\sim 2$ by~\citealt{Daddi2007ApJ} (black line, labeled `$\times 20 - {\rm MS}$'). A choice of $\mathcal{N}_{\rm SF}\geq 1$
    (blue lines) would intersect the 
    main sequence, while $\mathcal{N}_{\rm SF}\geq 0.1$
    (red lines) would preclude only very quiescent `dead' galaxies from being classified as starbursts in our model.}}
    \label{fig:sb_sep_fig}
\end{figure}

SFGs undergo star-formation in either a `main sequence' mode or a `starburst' mode~\citep{Rodighiero2011ApJ}. In the main sequence mode, star-formation is extended throughout the galaxy, and persists over timescales of 1-2 Gyr. In the starburst mode, activity is instead more centrally-concentrated, and substantially more intense. Criteria to differentiate between galaxies in the two modes are typically observationally-determined, according to their location in $U-V$ vs. $V-J$ colour-colour space~\citep[see][]{Williams2009ApJ}. However, alternative, physically-motivated means of distinguishing between starburst and main sequence SFGs have also been considered. 
For example, a simple 
cut-off based on galaxy sSFR may be used~\citep[e.g.][]{Merlin2018MNRAS, Girelli2019AA}, which may also be redshift-dependent~\citep[e.g.][]{Pacifici2016ApJ}. 
The physical interpretation of this criteria (other than an indication of the current level of `mass-normalised' star-formation in a galaxy) is, however, somewhat unclear and does not strictly capture the activity of a galaxy compared to its previous star-formation intensity, as may be useful for a means of defining systems undergoing a starburst episode. Thus, we consider a dimensionless normalised star-formation rate~\citep[introduced in][]{Carnall2018MNRAS}, which is the ongoing star-formation rate of a galaxy compared to its lifetime-average, i.e.:
\begin{equation}
    \mathcal{N}_{\rm SF}(t) = \mathcal{R}_{\rm SF}(t)/\langle \mathcal{R}_{\rm SF} \rangle_t \approx t \;\! \mathcal{R}_{\rm SF}(t) / M^{\star} \ .
    \label{eq:separation_criteria}
\end{equation}
Here we approximate the total stellar mass formed over the lifetime of a galaxy by its living stellar mass $M^{\star}$, and its (maximum) age by that of the Universe at the redshift it is located. {We adopt a threshold of $\mathcal{N}_{\rm SF}\geq 10$ to define starburst SFGs in our models, with the other SFGs being regarded as main sequence. In Fig.~\ref{fig:sb_sep_fig}, we show how the choice of this parameter affects the distinction of starburst and main sequence SFGs at different redshifts in the $\mathcal{R}_{\rm SF}-M^{\star}$ plane, where SFGs located above the line are classified as starbursts. Our adopted criteria sets starbursts {as galaxies with} around 20 times 
{the star-formation rate of a main sequence star-forming galaxy at fixed stellar mass (for the main sequence} as defined by~\citealt{Daddi2007ApJ} for $z\sim 2$ galaxies; see also~\citealt{Rodighiero2011ApJ}), thus selecting only the `outlier' 
population known to exist away from the 
$\mathcal{R}_{\rm SF}-M^{\star}$ relation describing the main sequence~\citep{Elbaz2007AA, Peng2010ApJ}. Practically, this criteria defines main sequence SFGs as those which have experienced a prior episode of star-formation, with all other SFGs being classified as starbursts undergoing star-formation at a rate that is at least 10 times that of their lifetime-average.}\footnote{{Of those galaxies detected in $\gamma$-rays listed in~\citet{Ajello2020ApJ_SFG}, our criteria classifies the LMC, SMC, M31 and M33 as non-starbursts. Some systems which may be colloquially regarded as starbursts due to certain regions of elevated star-forming activity, are however also not selected (e.g. NGC 253, with a starburst core, or M82 with its star-forming core but relatively quiescent disc). We consider this separation is more interesting, as it draws a distinction between the $\gamma$-ray emission from strong starbursts compared to systems with only core or local regional starburst activities which are overall more quiescent.}}

\section{Results}
\label{sec:results}

\subsection{EGB spectrum}
\label{sec:egb_spectrum}

\begin{figure}
    \centering
    \includegraphics[width=\columnwidth]{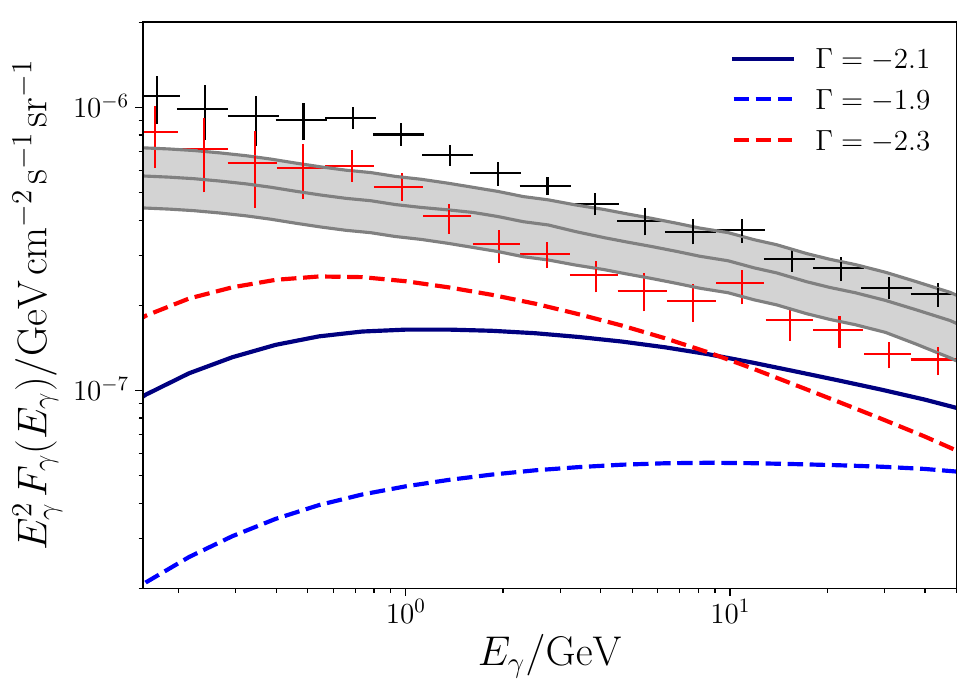}
    \caption{Fiducial model EGB spectrum between 0.1 and 50 GeV (navy line), 
    compared to the contribution from resolved 
    and unresolved blazars (grey band, denoting the three models of~\citealt{Ajello2015ApJ})
    {and the observed total EGB with 50 months of \textit{Fermi}-LAT data, shown in black, and the isotropic EGB, shown in red,  both adapted from~\citep[][as obtained using their foreground model A]{Ackermann2015ApJ}.} 
    The impact on the EGB spectrum by alternative choices of the CR spectral index is shown (red and blue dashed lines). These are representative of the range of values observed in nearby SFGs.}
    \label{fig:spec_plaw}
\end{figure}

The strongest SFG contribution to the EGB arises between 0.1 GeV and a few tens of GeV. This is shown by the navy line in Fig.~\ref{fig:spec_plaw} for our fiducial model.  
The contribution from resolved and unresolved blazars is indicated by the grey band (denoting the three models presented in~\citealt{Ajello2015ApJ}), {and data is shown for the total EGB spectrum with 50 months of \textit{Fermi}-LAT data, and the isotropic EGB component (see~\citealt{Ajello2015ApJ}; originally from~\citealt{Ackermann2015ApJ}) which indicate that 
the predicted EGB from our fiducial model does not exceed observational constraints, and can form a substantial component of the isotropic EGB over the energy range considered.} 
We also find good agreement between our fiducial model and results of other, recent works (in particular, that of~\citealt{Sudoh2018PASJ} and~\citealt{Peretti2020MNRAS}; see Appendix~\ref{sec:comparison_other_work} for details). The results of~\cite{Roth2021Natur} yield a higher EGB intensity, which is sufficient to account for the entire unresolved extragalactic $\gamma$-ray emission. We consider their higher value to stem primarily from differences in their treatment of CR transport within SFGs compared to this work (for discussion, see Appendix~\ref{sec:comparison_other_work}).

The effect of variations in the {injected} CR spectral index adopted for the SFG population is also shown in Fig.~\ref{fig:spec_plaw}. The choices of $\Gamma=-1.9$ and $\Gamma=-2.3$ reflect the range of values inferred for nearby starburst galaxies~\citep{Ajello2020ApJ_SFG}. As considered in~\cite{Ambrosone2021MNRAS}, these alternative spectral index values can have a noticeable impact on the predicted shape and intensity of the resulting EGB spectrum. However, we note that these choices are at the extremes of the observed distribution of CR indices, with the mean being closer to our fiducial choice. Accounting for spectral index variations by modelling $\Gamma$ with a plausible distribution of values blended over a SFG population would therefore not be likely 
to differ greatly from our fixed-value result. 

\begin{figure}
    \centering
    \includegraphics[width=\columnwidth]{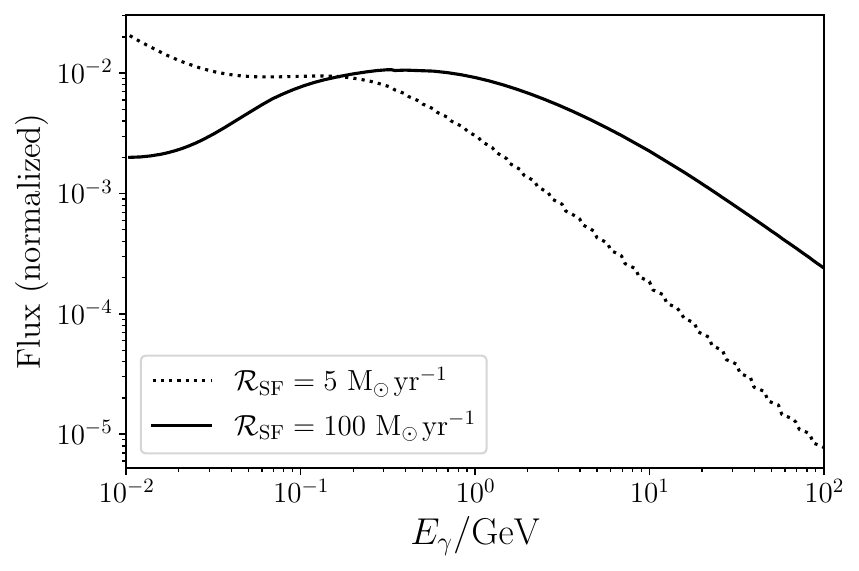}
    \caption{{Normalized $\gamma$-ray flux $E_{\gamma} \;\! F(E_{\gamma})$ for two galaxies at $z=0$, with stellar masses $M^{\star} = 10^9 \;\!{\rm M}_{\odot}$, with star-formation rates of $\mathcal{R}_{\rm SF} = 5 \;\! {\rm M}_{\odot}\;\!{\rm yr}^{-1}$ and $\mathcal{R}_{\rm SF} = 100 \;\! {\rm M}_{\odot}\;\!{\rm yr}^{-1}$, as labeled. Spectra are normalized to unity, to allow comparison between spectral shapes, where the starburst SFG presents a noticeably harder spectrum.}}
    \label{fig:sb_qui_compare}
\end{figure}

{Although we adopt a fixed spectral index for the injection of CRs in our model, we note that the steady-state CR spectrum for SFGs can differ from this based on the galaxy model parameters. This is due to our energy-dependent diffusion and pp attenuation treatment of the hadronic CRs (cf. equation~\ref{eq:cr_spec1}), which can settle into a softer spectrum in more quiescent galaxies. This is illustrated by the resulting $\gamma$-ray emission of two example SFGs in Fig.~\ref{fig:sb_qui_compare}, where the starburst SFG with $\mathcal{R}_{\rm SF} = 100 \;\! {\rm M}_{\odot}\;\!{\rm yr}^{-1}$ presents a substantially harder $\gamma$-ray spectrum than the less active $\mathcal{R}_{\rm SF} = 5 \;\! {\rm M}_{\odot}\;\!{\rm yr}^{-1}$ galaxy. This reflects differences seen between $\gamma$-ray emission spectra of starburst and more quiescent galaxies~\citep[e.g.][]{Tamborra2014JCAP}.}

\subsubsection{Contribution from SFG types}
\label{sec:SFG_mass_contributions}

\begin{figure}
    \centering
    \includegraphics[width=\columnwidth]{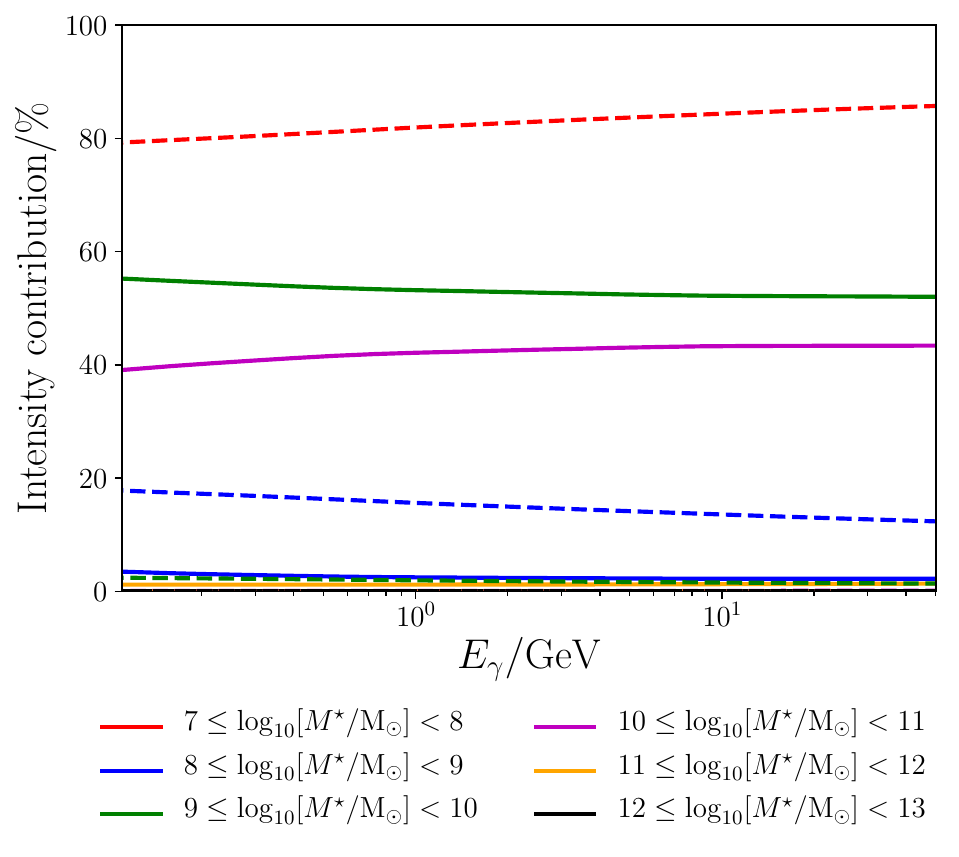}
    \caption{{Relative contribution of different classifications of SFGs 
    over the spectral range shown in Fig.~\ref{fig:spec_plaw}, separated according 
    to galaxy stellar mass in six logarithmic bands as indicated. \textbf{Solid lines} show the contribution from main sequence SFGs. \textbf{Dashed lines} show the contribution from starburst SFGs. Overall, the total contribution (not shown) is strongly dominated by starbursts.}
    }
    \label{fig:flat_spec}
\end{figure}

Fig.~\ref{fig:flat_spec} shows the relative EGB contribution of different 
components of the SFG population, 
split according to galaxy stellar mass and star-formation mode.
Main sequence and starburst SFGs are classified according to the criteria in section~\ref{sec:sb_and_ms} and, although a substantial fraction of the SFG population resides in the main sequence, the EGB contribution is strongly dominated by SFGs in the starburst mode. 
This is because our separation criteria selects starburst SFGs according to characteristics which also tend to make them
 effective sources of $\gamma$-rays.\footnote{Given the dependence of 
the molecular gas density on $\mathcal{R}_{\rm SF}$ (rather than stellar mass), 
and that this molecular gas forms the primary $\gamma$-ray emission volume for a SFG, 
it follows that the starburst/main sequence separation criteria in this work is also an effective 
means of identifying strongly $\gamma$-ray emitting populations of SFGs.} 
As such, they far out-shine the main sequence SFGs, and provide 
{93.7 percent of the SFG contribution to the EGB in our model at 0.01 GeV, rising to 99.7 percent by 50 GeV.} 

Although other studies~\citep[e.g.][]{Sudoh2018PASJ} also find starburst SFGs to be an important $\gamma$-ray source population in the EGB, 
their contribution was not found to dominate the emission from all SFGs. 
We consider this to be due to differences in starburst/main sequence population separation criteria adopted. In~\citealt{Sudoh2018PASJ}, for example,
sufficient information was available 
in their 
input semi-analytic model (the \textit{Mitaka} model -- see~\citealt{Nagashima2004ApJ})
to explicitly distinguish galaxies experiencing intense star-formation after a major merger event. 
By contrast, in the present work, the separation criteria (section~\ref{sec:sb_and_ms}) is instead based on the galaxy physical properties. 
The requirement for a high star-formation rate compared to the average $\mathcal{R}_{\rm SF}$ over a galaxy's lifetime equivalently selects for systems with a high star-formation rate for low stellar mass (cf. equation~\ref{eq:separation_criteria}), which is not explicitly required by the~\citet{Sudoh2018PASJ} criteria.

Fig.~\ref{fig:flat_spec} further shows that the EGB intensity in our model comes predominately from starburst SFGs of relatively low stellar masses (over 80\% of the EGB intensity comes from SFGs of stellar mass $M^{\star} = 10^{7}-10^{8}\;\!{\rm M}_{\odot}$, and less than 5\% originates in galaxies of $M^{\star} >10^{9}\;\!{\rm M}_{\odot}$), and this is largely independent of $\gamma$-ray energy. 
By contrast, the $\gamma$-ray contribution from main sequence SFGs primarily comes from those with higher stellar masses (over 99\% of the main sequence contribution is from galaxies of stellar mass {$M^{\star} = 10^{9}-10^{11}\;\!{\rm M}_{\odot}$}). 
As the main sequence contribution to the overall emission is negligible, it can thus be considered that the EGB is a biased tracer of CR interactions and (hence) feedback activity in starburst SFGs of low stellar mass. Our results suggest that the EGB would be relatively unhelpful to yield information about CR processes in distant populations of more massive, developed, disk-like galaxies, such as those 
detected in $\gamma$-rays in the local Universe (e.g. M82).

\subsection{SFG contributions over redshift}
\label{sec:contribution_over_z}

While the EGB spectrum is the total intensity for the $\gamma$-ray emission from SFGs integrated over cosmic time, the evolution of those sources can also be probed by considering the redshift from which the EGB $\gamma$-rays originated. This can provide insight into which epochs in cosmic history are contributing most strongly, and indicate observable tracers that can be used to track the evolution of the underlying sources. 

\begin{figure}
    \centering
    \includegraphics[width=\columnwidth]{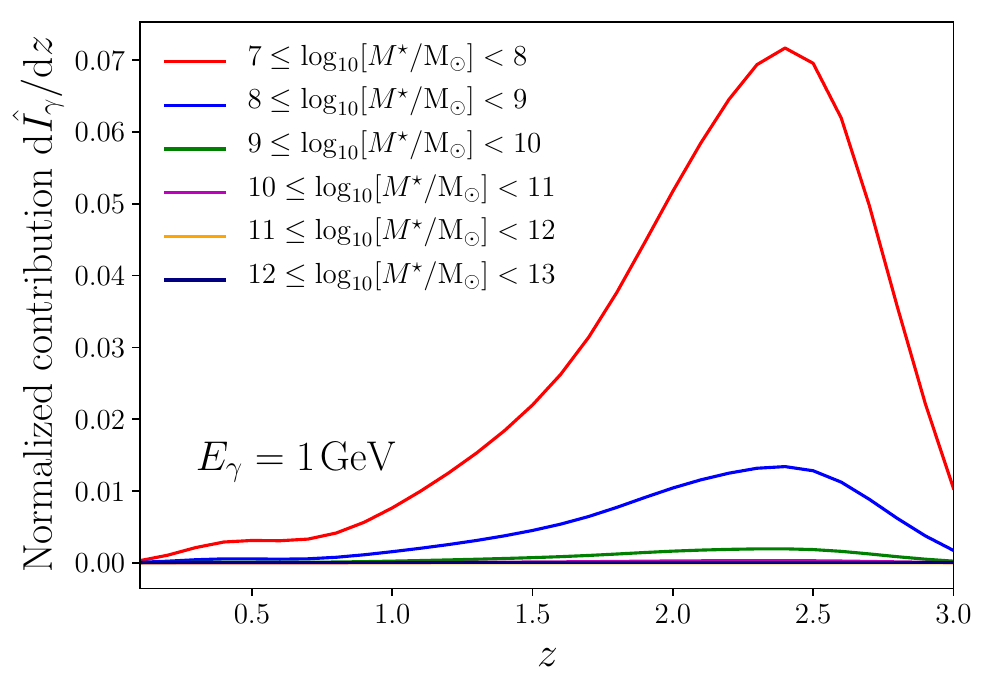}
    \caption{Intensity contribution to EGB at $E_{\gamma} = 1$ GeV from mass-separated SFGs over redshift, normalised to the total SFG EGB intensity at 1 GeV. The 
    majority of the emission originates from low mass galaxies, $M^{\star} = 10^7 - 10^8 \;\!{\rm M}_{\odot}$, peaking at around $z\sim 2.5$. A version of this plot normalised to the total emission in each mass band is shown in panel (c) of Fig.~\ref{fig:normed_energies}, where the differences in redshift evolutions can be seen more clearly. }
    \label{fig:normed_1GeV_z}
\end{figure}

{The SFG origin of the EGB varies over redshift. This can be seen in 
Fig.~\ref{fig:normed_1GeV_z}, where the EGB emission distribution from SFG populations over redshift is shown.  To aid comparison between galaxy mass bands, contributions from galaxies of different masses are 
normalised by the total $z=0$ EGB intensity at 1 GeV.} As previously shown by Fig.~\ref{fig:flat_spec}, the majority of the emission originates from low mass galaxies, $M^{\star} = 10^7 - 10^8 \;\!{\rm M}_{\odot}$, however Fig.~\ref{fig:normed_1GeV_z} further reveals that the bulk of the GeV emission comes from SFGs located between $z\sim 2$ and 2.5. Around 15\% of the total emission is contributed by slightly higher mass galaxies, around $M^{\star} = 10^8 - 10^{9} \;\!{\rm M}_{\odot}$, and it can be seen that the peak of their contribution originates from a slightly {lower} redshift, around {$z\sim 2.4$}, with the trend to {marginally lower peak} redshifts continuing with increased stellar mass. 
{This demonstrates that the} EGB does not, therefore, reliably trace cosmic star-formation history (as may otherwise be assumed, given the relation between CR abundance and $\mathcal{R}_{\rm SF}$), which peaks at around $z\sim 2$~\citep[e.g.][]{Madau2014ARAA}, and indicates that the SFG origin of the EGB offers a window to view CR processes in galaxies around 0.7 Gyr before the cosmic noon. 

\begin{figure*}
    \centering
    \includegraphics[width=0.9\textwidth]{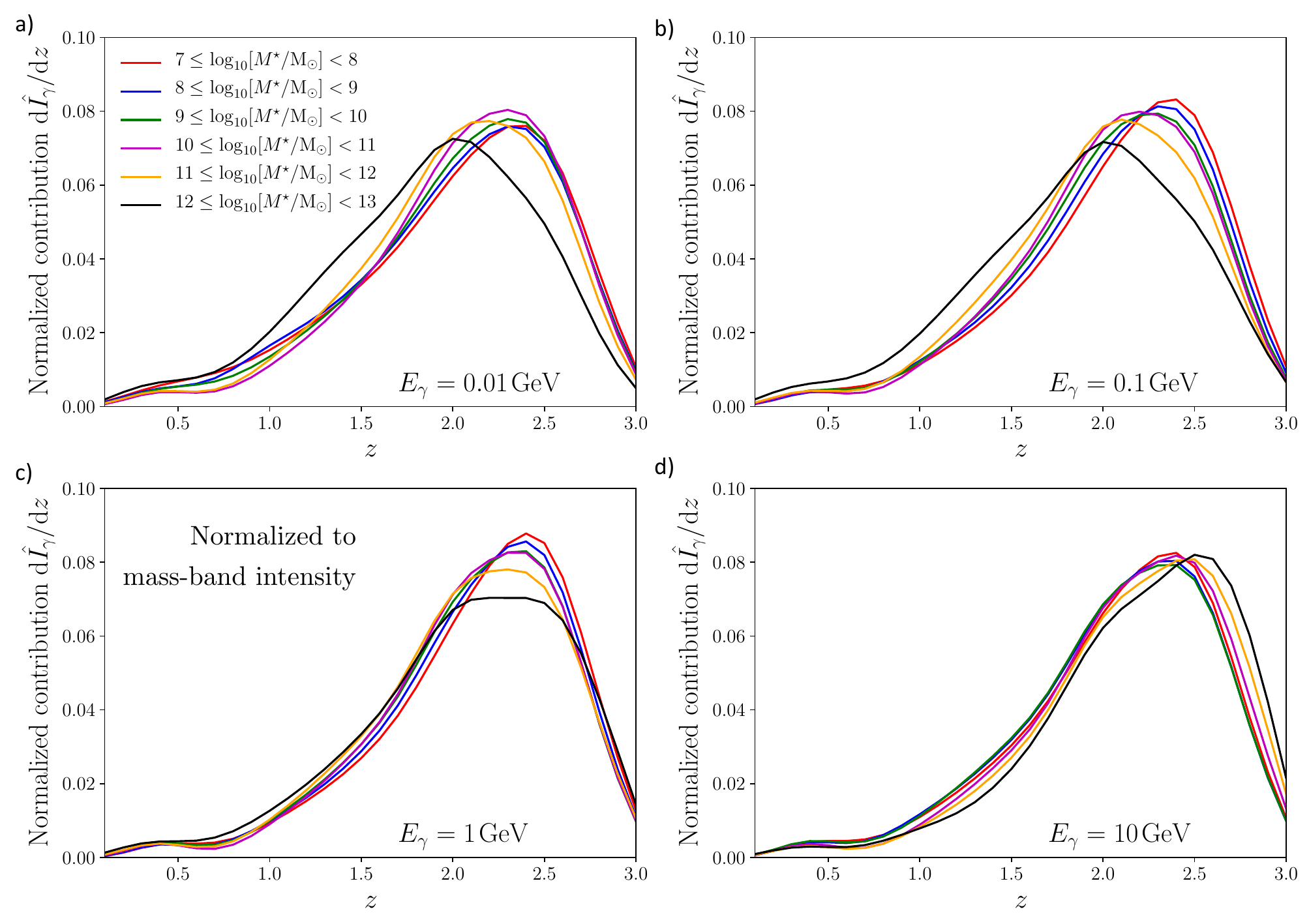}
    \caption{Redshift distribution of the $\gamma$-ray contribution to the EGB from all SFGs (main sequence and starburst), at $z=0$ $\gamma$-ray energies of 0.01, 0.1, 1 and 10 GeV in panels (a), (b), (c) and (d), respectively. Distributions are separated according to galaxy stellar masses, and normalised to the total $z=0$ $\gamma$-ray intensity in each mass band (to aid comparisons between distributions). 
    This shows a noticeable difference in the peak of $\gamma$-ray emission with energy for the same source population, {reflecting the evolution of physical SFG properties that modify their $\gamma$-ray emission spectra.}}
    \label{fig:normed_energies}
\end{figure*}

The reason for this discrepancy between the peak redshift of $\gamma$-ray emission from SFGs and the peak in cosmic star-formation lies {in the redshift distribution of those galaxies contributing to the bulk of the emission.\footnote{{Additional factors, such as the intensity of ambient radiation fields (which can attenuate $\gamma$-rays at higher energies), or the evolution of the ability of a galaxy to contain CRs and the corresponding changes in the $\gamma$-ray emission spectrum with the galaxy properties (cf. Fig.~\ref{fig:sb_qui_compare}) could also have subsidiary impacts.}} Fig.~\ref{fig:normed_1GeV_z} shows that $\gamma$-rays are predominantly contributed to the EGB by lower mass galaxies. In the galaxy population model of~\cite{Katsianis2017MNRAS} underlying our results, the total star-formation activity in these lower mass galaxies (and hence their $\gamma$-ray emission) peaks at higher redshifts than in SFG populations overall. A consequence of this would be that the redshift distribution of the most strongly $\gamma$-ray emitting SFGs is biased compared to the cosmic star-formation history overall, with the `$\gamma$-ray noon' significantly preceding the  `star-formation noon'. }

This effect is also evident in Fig.~\ref{fig:normed_energies}, where we model the total SFG contribution to the EGB at different energies, $E_{\gamma} = 0.01$, 0.1, 1 and 10 GeV. Compared to Fig.~\ref{fig:normed_1GeV_z}, the distributions are now normalised to the total intensity in each band of stellar mass, rather than the total SFG $\gamma$-ray intensity. This is to aid comparison between the redshift distributions of different galaxy mass classes. In all cases, the lowest mass SFGs, between $M^{\star} = 10^7 - 10^8\;\!{\rm M}_{\odot}$, dominate the emission (cf. Fig.~\ref{fig:flat_spec}). 
Although the redshift distribution of $\gamma$-ray emission at the four energies considered in Fig.~\ref{fig:normed_energies} is different, the underlying SFG population is the same. 
{This follows from the spectral variation of SFGs with different star-formation activity (cf. Fig.~\ref{fig:sb_qui_compare}). Those SFGs with higher star-formation rates present harder spectra, suggesting that the proportion of $\gamma$-rays at higher energies would be boosted during the cosmic noon. This can account for the broader evolutionary peaks seen at 1 GeV, and the $z\sim 2$ `bump' emerging in all galaxy mass bands at 10 GeV.}

Fig.~\ref{fig:normed_energies} also shows that higher mass SFGs present different evolutionary trends in their $\gamma$-ray emission to their lower mass counterparts, especially at lower $\gamma$-ray energies. At 0.01 GeV and 0.1 GeV, the evolution of $M^{\star} = 10^{12} - 10^{13}\;\!{\rm M}_{\odot}$ galaxies is much closer to the underlying redshift distribution of the SFG source population (which is broadly reflective of the cosmic star-formation history, peaking at $z\sim 2$; see~\citealt{Katsianis2017MNRAS}, and~\citetalias{Owen2021MNRAS} for discussion). Below a few GeV, we find a significant component of the $\gamma$-rays from a SFG is driven by leptons. While both primary and secondary electrons contribute to this, the emission component from secondary electrons typically becomes more important than that from primary electrons above $\sim$ 1 GeV. The production of secondary electrons is mediated by the pp interaction, and would be subject to the same processes that mediate pion-decay $\gamma$-ray emission from intensely star-forming galaxies. This does not apply to the primary electrons, whose abundance (and $\gamma$-ray emission) would be directly proportional to a galaxy's star-formation rate. {This competition between primary and secondary leptonic $\gamma$-ray emission is particularly evident among higher mass SFGs as, in our model, the radius of a SFG is specified by its stellar mass (cf.  equation~\ref{eq:size_mass_relation}). More massive galaxies would typically be larger, and 
this would ensure more complete containment of CR electrons - in particular, primary electrons. Secondary electrons are injected by protons, which are less confined (since they cool more slowly, protons have more time to undergo diffusive escape, for example). As such, in more massive galaxies, primary leptonic $\gamma$-ray emission and would dominate over the secondary electron contribution up to higher energies than in lower-mass systems. This primary leptonic emission more closely traces the star-formation rate of a galaxy, particularly at low $\gamma$-ray energies, and accounts for the different behaviour of the high-mass SFG band up to 1 GeV in Fig.~\ref{fig:normed_energies}.}

\subsubsection{{Comparison with previous work}}

{The peak EGB $\gamma$-ray emission at $z\sim 2.5$ recovered by our model differs substantially to that reported in previous studies. In particular, ~\cite{Makiya2011ApJ},~\cite{Tamborra2014JCAP} and~\cite{Roth2021Natur} each found a peak EGB contribution originating from $z\sim 1$. In all cases, this discrepancy  appears to stem mainly from differences in the underlying redshift distribution of sources used in our model compared to the previous works. \cite{Makiya2011ApJ} introduced a model where the $\gamma$-ray emission of their SFG sources was obtained from the product of their star-formation rate and gas mass. This was applied to a source population obtained from a semi-analytic model of hierarchical galaxy formation (the \textit{Mitaka} model -- see~\citealt{Nagashima2004ApJ}; also used in the work of~\citealt{Sudoh2018PASJ}) to model their EGB contribution. 
The redshift distribution of these sources (and, indeed, the product of their star-formation rate and gas mass used to determine the SFG $\gamma$-ray luminosity) was shown to peak  around $z\sim 1$, which was then inherited by their EGB model. Similarly, \cite{Tamborra2014JCAP} based their EGB model on a redshift distribution of SFGs that peaked at $z\sim 1$~\cite[obtained from][]{Gruppioni2013MNRAS}, with the $\gamma$-ray emission redshift distribution following from this. 
}

{\cite{Roth2021Natur} 
used the Cosmic Assembly
Near-infrared Deep Extragalactic Legacy Survey in the
Great Observatories Origins Deep Survey S field~\citep{Grogin2011ApJS} to inform their $\gamma$-ray SFG source population, including inferred galaxy structural parameters~\citep{Wel2012ApJS}. This shows a lower peak in its redshift distribution than considered in our model. Moreover, the \cite{Roth2021Natur} model invoked a more detailed treatment of CR containment within SFGs, and their computed calorimetric
fraction shows an increase around z$\sim$1 (this contrasts with the uniform CR containment fraction adopted in our model). This leads to their lower recovered redshift peak in $\gamma$-ray emission.}

\subsection{Small scale anisotropy signatures}

\begin{figure*}
    \centering
    \includegraphics[width=0.9\textwidth]{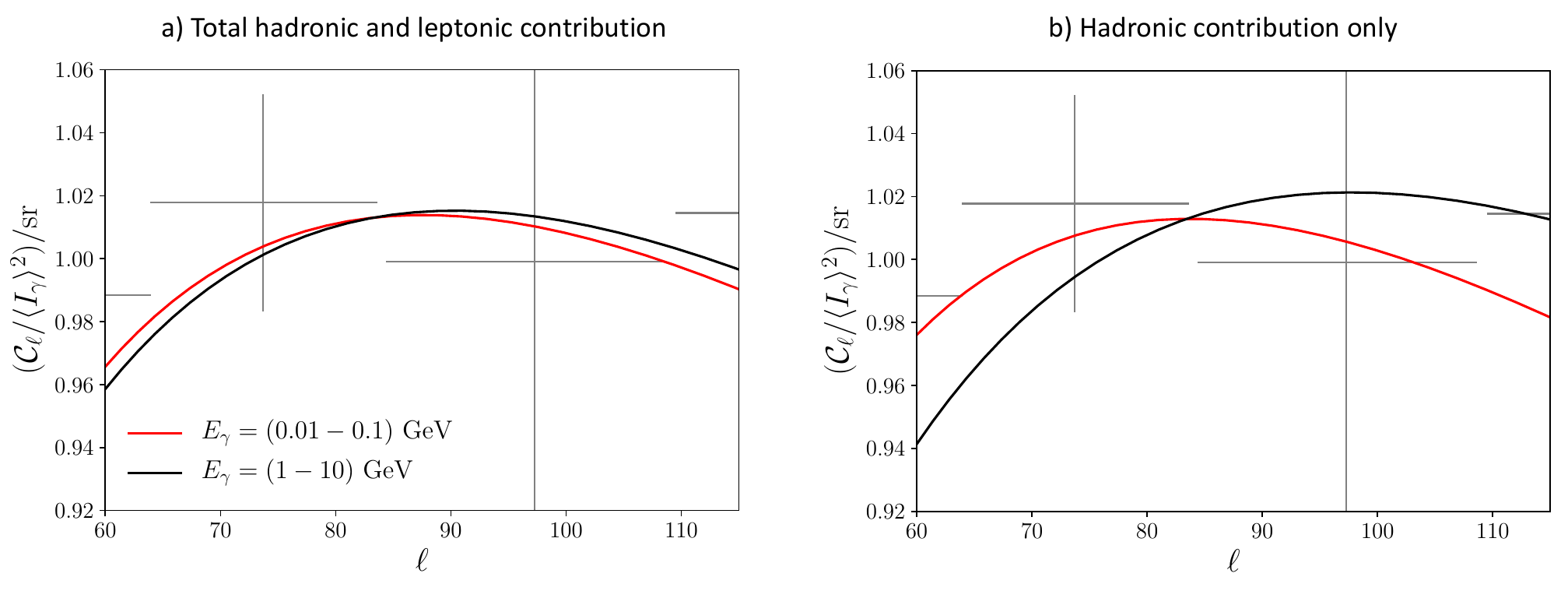}
    \caption{Small-scale anisotropy power spectrum $\mathcal{C}_{\ell}$s shown between multipoles of $\ell=60$ and $\ell = 120$
    for the fiducial model, {with panel (a) showing the result for the total diffuse EGB, and panel (b) showing the limit without leptonic $\gamma$-ray emission}. In both panels, anisotropy at different energies are compared. 
    $\mathcal{C}_{\ell}$s are normalised to the average EGB intensity in their respective energy band. 
    {Data obtained with \textit{Fermi}-LAT (adapted from~\citealt{Fornasa2016PhRvD}) is plotted for the energy band (1.38-1.99) GeV for comparison, where 3FGL sources have been masked. This shows our model to be approximately consistent with observations.}}
    \label{fig:cls}
\end{figure*}

As described in section~\ref{sec:anisotropies_egb}, source classes in the EGB with different redshift distributions would imprint small-scale anisotropies at different angular scales. Analysis of the anisotropic power spectrum of the EGB can thus reveal these signatures. Fig.~\ref{fig:normed_energies} demonstrated that the imprinted redshift distribution of EGB $\gamma$-rays from SFGs is different for different energies due to the intrinsic evolution of the physical properties of the sources, and this would accordingly imprint small scale anisotropies at correspondingly different scales. This can be seen in Fig.~\ref{fig:cls},
where anisotropy signatures in two different $\gamma$-ray energy bands are shown, normalised to the average EGB intensity in the respective band. {Panel (a) shows the results from our fiducial model. For comparison, we additionally show the case where leptonic emission is removed from the fiducial model in panel (b), i.e. neutral pion-decay emission only. \textit{Fermi}-LAT data shown for comparison are adapted from~\cite{Fornasa2016PhRvD}, for the diffuse EGB in the energy band (1.38-1.99) GeV. }

 {In the fiducial case (Fig.~\ref{fig:cls}, panel a), there is only marginal difference between the anisotropy signatures imprinted in the two energy bands. Both show a peak at around $\ell \sim 90$, which originates from the dominant SFG source population of $M^{\star} = 10^7 - 10^8\;\!{\rm M}_{\odot}$. In Fig.~\ref{fig:normed_energies} these can be seen to follow a similar redshift distribution at all energies, with a peak around $z\sim 2.5$. The leptonic component of our model contributes strongly to the EGB emission considered here, particularly in the lower energy band. However, this is subject to uncertainties due to faster cooling times than hadrons, and greater sensitivity to the sub-galactic variations of conditions in SFGs that were beyond the scope of this work. As such, there is value in considering an alternative scenario, where leptonic emission from SFGs is significantly lower than found by our fiducial model. Panel (b) of Fig.~\ref{fig:cls} considers the extreme limit of this, where only hadronic $\pi^0$ decay contributes to the EGB (and reflects the SFG prototype emission model invoked previously, in~\citetalias{Owen2021MNRAS}). Here, a much greater difference emerges in the EGB anisotropy in the two energy bands. 
 The lower energy band $E_{\gamma} = (0.01 - 0.1) \;\!{\rm GeV}$, now shows stronger anisotropy at larger scales, with a peak around $\ell \sim 100$. By contrast, the higher energy band, $E_{\gamma} = (1 - 10) \;\!{\rm GeV}$ shows a broader distribution of power (a less pronounced peak), with the strongest anisotropy signature at smaller scales, around $\ell \sim 80$. This reflects the differences in the underlying redshift distribution 
 that are practically `hidden' by leptonic emission in our fiducial model and suggests that, if leptonic $\gamma$-ray emission is lower than considered by our model, EGB anisotropies at different energies may reveal evidence of the evolving physical conditions of SFG source populations.}

\subsection{Discussion and remarks}

\subsubsection{The EGB as a biased tracer for CR feedback in galaxies}

$\gamma$-ray emission traces {CR} interactions, and therefore has potential as a proxy to infer the action of CR feedback operating within a galaxy. The EGB thus offers potential to investigate the role CRs have in shaping the evolution of distant populations galaxies, and their action as feedback agents that may have helped to bring about the end of star-formation after the cosmic noon. 
However, this work has shown that careful consideration of the physical nature of the underlying source populations is required to do this reliably. Even with the relatively modest amount of modification to $\gamma$-ray emission spectra that can arise at GeV energies due to varying physical conditions in source galaxies, 
substantial biases can be introduced to the inferred redshift distribution of those sources as might be determined by EGB anisotropy analyses (cf. Fig.~\ref{fig:normed_energies}). 
Moreover, 
this work has shown that the vast majority of the GeV emission from SFGs in the EGB is actually contributed by 
low mass starbursts (cf. Fig.~\ref{fig:flat_spec}), and these are mainly located \textit{before} the noon of cosmic star-formation (cf. Fig.~\ref{fig:normed_1GeV_z}).
While this does not prevent the EGB from being used to 
constrain CR processes in such systems, 
it is important to consider that the EGB would predominantly be providing information about how CRs engage with young, low mass SFG populations - and this may be quite different to the role they play in higher mass systems that cannot easily be accessed using analyses of the $\gamma$-ray background.

\subsubsection{Considerations at higher energies}

EGB anisotropy at energies below a few 10s of GeV, as shown in Figure~\ref{fig:cls} are not significantly affected by EBL reprocessing effects, and are only moderately affected by internal $\gamma$-ray attenuating processes. 
However, the opposite is true at higher energies (e.g. TeV with up-coming facilities like the Cherenkov Telescope Array, CTA - see~\citealt{CTA2019_book}), where 
analysis of EGB anisotropies would open up new opportunities to probe the physical conditions in distant SFGs much more rigorously. In addition to a greater importance of EBL cascades in modifying EGB signatures, the TeV $\gamma$-ray emission from a SFG is much more strongly affected by attenuation and reprocessing in radiation fields associated with both stars and interstellar dust, and would be influenced by both stellar population properties and physical dust characteristics. Moreover, as attenuation effects are more severe, TeV EGB signatures would also become more sensitive to properties of the interstellar medium of host galaxies, in particular their density configurations, molecular gas clumpiness/morphology, and stellar distributions. Together these would determine how brightly SFGs would emit TeV $\gamma$-rays, and much more variation between systems would likely arise.  

Future studies using the TeV $\gamma$-ray background would depend on substantially more detailed treatments of the physical properties of SFG populations and their internal conditions than has been necessary at GeV energies. In particular, more sophisticated models of the 
intensity, spatial geometry and temperatures of stellar and dust-reprocessed radiation fields in SFGs would be required, 
including their dependence on global galaxy properties provided by population models (for example redshift, stellar mass and/or star-formation rate -- see, e.g. ~\citealt{Liang2019MNRAS}) and stellar distribution/obscuration and internal attenuation patterns ~\citep[e.g.][]{Lin2021MNRAS}. 
Such developments 
are already possible, and can be informed by multi-wavelength data with spectral fitting~\citep[e.g.][]{Kim2021MNRAS} in a range of galaxy types 
to allow reasonable estimates and/or scaling relations to be constructed for energy-dependent $\gamma$-ray escape fractions of galaxies with broad ranges of physical properties and internal configurations. 

\section{Conclusions}
\label{sec:conclusions}

In this work, we investigated the contribution of SFGs to the EGB at energies between 0.01 GeV and 50 GeV. We used a physically-motivated SFG $\gamma$-ray prototype model to compute $\gamma$-ray emission spectra from galaxy populations based on their redshift, star-formation rate and physical properties. In this model, the $\gamma$-ray emission was driven by 
hadronic {and leptonic} interactions of CRs, and the emitted spectrum was modified by attenuating energy-dependent pair-production processes in interstellar radiation fields. This is a refinement of the model previously introduced in~\citetalias{Owen2021MNRAS}, in which we now put more detailed focus on the impacts of variations of physical galaxy properties in modifying their $\gamma$-ray luminosity and spectrum. 

We computed an EGB spectrum by applying our prototype SFG to a galaxy population model, 
and found this to be broadly consistent with previous works~\citep[e.g.][]{Sudoh2018PASJ, Peretti2020MNRAS}, although our predicted EGB intensity was lower than that recently found by~\cite{Roth2021Natur} due to differences in the treatment of CR propagation and escape from the SFGs. We considered the EGB contributions from sub-populations of SFGs, split firstly according to their mode of star-formation, then secondly according to their masses. We found that starburst SFGs (specified as those galaxies experiencing star-formation at a rate of at least 10 {times that} of their lifetime average) completely dominate their EGB contribution, with the contribution from main sequence SFGs being practically negligible -- however this conclusion is dependent on the criteria used to select starbursts, and we consider that alternative choices leading to different results would be no less valid. We also found that low mass starbursts (of stellar masses between $M^{\star} = 10^7 - 10^8 \;\!{\rm M}_{\odot}$) are responsible for the majority of the SFG contribution to the EGB, accounting for more than 80\% of the EGB intensity in our model. 

{We showed that the EGB spectrum at different energies is sensitive to starburst SFGs of low stellar mass and that, at most energies, the emission contributing to the diffuse EGB is dominated by galaxies around $z\sim 2.5$, several hundred Myr {before} the cosmic noon. We showed how different populations of galaxies would imprint small-scale anisotropy signatures in the EGB at different angular scales based on their redshift distribution, allowing their contributions to be resolved using spatial EGB analyses, and how these anisotropy signatures could change depending on the balance of CR protons to electrons in SFGs. We further showed how these small-scale anisotropies could 
provide a means to indirectly 
probe the changing interstellar radiation fields and molecular gas abundances of low-mass galaxies before the high noon of cosmic star-formation, and would open up a valuable new way to assess the role played by CRs in shaping the evolution of galaxies.} 

\section*{Data Availability}

No new data were generated or analysed in support of this research.

\section*{acknowledgements}

This work used high-performance computing facilities operated by the
Center for Informatics and Computation in Astronomy (CICA) at National
Tsing Hua University (NTHU). This equipment was funded by the Ministry of
Education of Taiwan and the Ministry of Science and Technology (MOST) of Taiwan.
We thank the National Center for High-performance Computing, Taiwan, for providing computational and storage resources, and the National Center for Theoretical Sciences, Taiwan, for provision of HPC time allocation, supported by a grant from MOST (110-2124-M-002-012). 
ERO is supported by the Ministry of Education of Taiwan at CICA, NTHU. AKHK acknowledges support from MOST (grant 110-2628-M-007-005). KGL acknowledges support from JSPS Kakenhi Grants JP18H05868 and JP19K14755.  
This research used NASA's Astrophysics Data Systems. The authors thank the anonymous referee for their insightful comments, which substantially improved the manuscript. 



\bibliographystyle{mnras} 
\bibliography{references} 



\appendix

\section{Comparison with previous works}
\label{sec:comparison_other_work}

\begin{figure*}
    \centering
    \includegraphics[width=\textwidth]{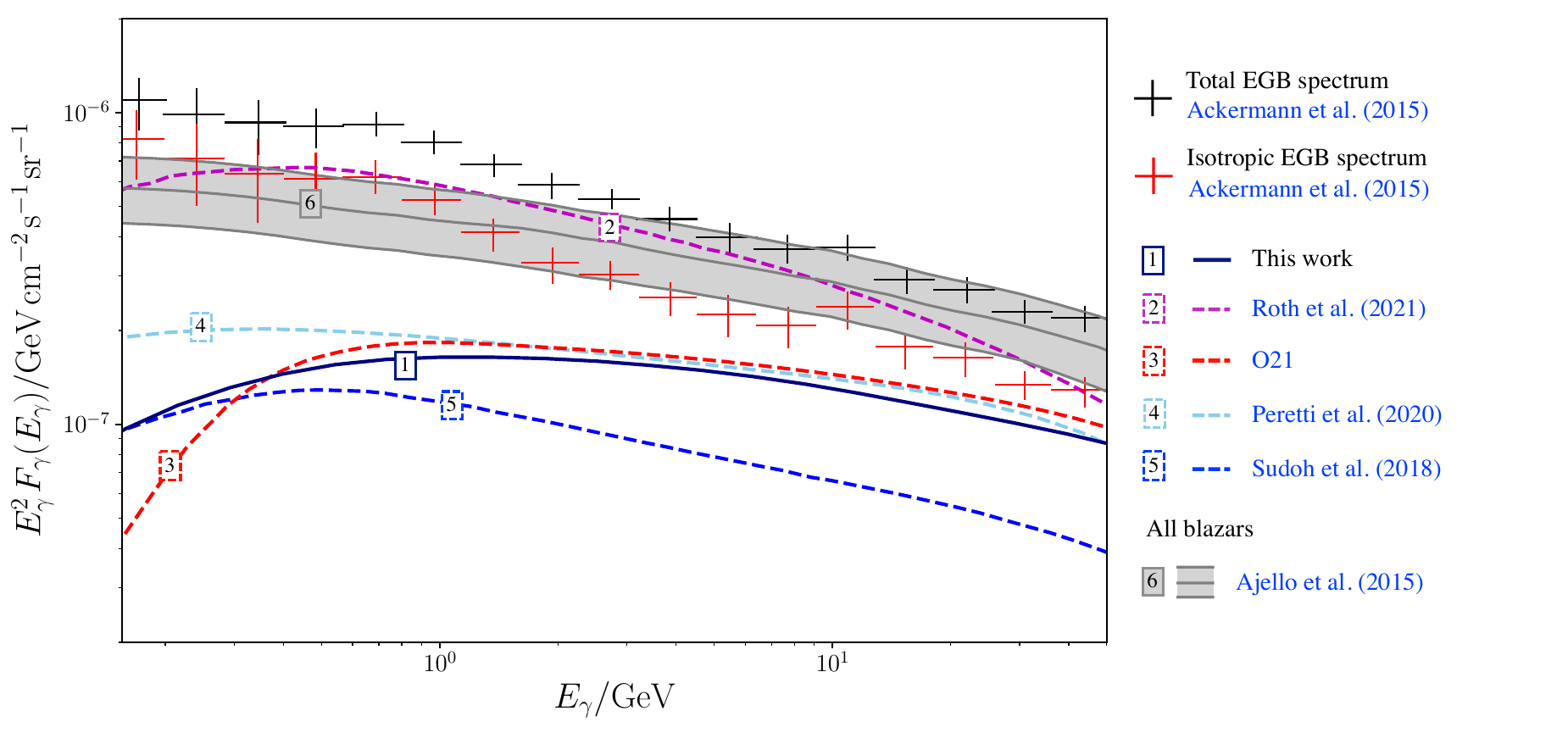}
    \caption{Total contribution from SFGs (starburst and main sequence) to the {isotropic} EGB, between 0.1 and 50 GeV. The fiducial result from this work is given by line 1, which is in agreement with the constraint imparted by the  contribution from resolved 
    and unresolved blazars (grey band, denoting the three models of~\citealt{Ajello2015ApJ})
    and the observed EGB with 50 months of \textit{Fermi}-LAT data~\citep{Ackermann2015ApJ}, determined using their foreground model A. Comparison is made with four recent works; \citealt{Roth2021Natur} (line 2), \citetalias{Owen2021MNRAS} (line 3), \citealt{Peretti2020MNRAS} (line 4), and~\citealt{Sudoh2018PASJ} (line 5).}
    \label{fig:spectrum_compare_appdx}
\end{figure*}

Comparison between the total {isotropic} EGB spectrum given by our fiducial model and other works is shown in Figure~\ref{fig:spectrum_compare_appdx}.  Here, it can be seen that the model EGB spectrum of~\cite{Roth2021Natur} yields a substantially higher intensity, which is sufficient to account for the entire unresolved EGB. The principal difference between this work and~\citet{Roth2021Natur} is in our prototype model and its treatment of CR transport. In particular, in the present work, CR removal by advection is accounted for by a single reduction factor at all energies. 
While considered appropriate for this demonstrative study (and also the results of~\citetalias{Owen2021MNRAS}), future model refinements will include a more robust treatment of this CR transport physics and its variation according to SFG physical conditions, which can have implications for the predicted EGB~\citep{Ambrosone2022arXiv}. 

The fiducial model of~\citetalias{Owen2021MNRAS} arguably offers the closest comparison to the results of this paper, as many aspects of the model are identical. Indeed, this is reflected by the similarity between the predicted EGB spectra above 1 GeV. At lower energies, the more physical modelling of SFG environments adopted in this work, and the provision for their variation over redshift and with galaxy properties, leads to a higher EGB intensity compared to~\citetalias{Owen2021MNRAS}, with better consistency evident instead with~\cite{Sudoh2018PASJ} and~\cite{Peretti2020MNRAS}. 

The approach of~\cite{Peretti2020MNRAS} is fundamentally similar to this work, however their prototype model is based on the $\gamma$-ray emission spectrum of M82 and then scaled to other galaxies according to star-formation rate. This does not account for certain variations in physical galaxy properties that the present work put focus on, however it does {capture the same underlying $\gamma$-ray emission processes}. 

\cite{Sudoh2018PASJ} adopt a very different approach to that used here - and, indeed, to the other models presented in Fig.~\ref{fig:spectrum_compare_appdx}. Their EGB model is developed from a semi-analytic model of galaxy formation~\citep{Nagashima2004ApJ}. Despite this, the total {isotropic} EGB spectrum still shows good agreement with this work, particularly at lower energies.

\bsp	
\label{lastpage}
\end{document}